\documentclass[a4paper]{jpconf}
\usepackage{amsfonts, amsmath, amssymb, amsthm, amsbsy, amscd}

\theoremstyle{definition}
\newtheorem{example}{Example}
\newtheorem{counterEx}[example]{Counterexample}

\theoremstyle{remark}
\newtheorem{rem}{Remark}

\newcommand{\pinner}{\mathbin{\mathchoice
   {\hbox{\vrule width0.6em depth0pt height0.4pt
   \vrule width0.4pt depth0pt height0.8ex}}
   {\hbox{\vrule width0.6em depth0pt height0.4pt
   \vrule width0.4pt depth0pt height0.8ex}}
   {\hbox{\kern0.14em
   \vrule width0.48em depth0pt height0.4pt
   \vrule width0.4pt depth0pt height0.6ex\kern0.14em}}
   {\hbox{\kern0.1em
   \vrule width0.39em depth0pt height0.4pt
   \vrule width0.4pt depth0pt height0.5ex\kern0.1em}}}}

\DeclareMathOperator{\Jac}{Jac}
\DeclareMathOperator{\dvol}{d
vol}
\DeclareMathOperator{\supp}{supp}

\newcommand{\quasieq}{\mathrel{\text{``}{=}\text{''}}}
\newcommand{\quasicong}{\mathrel{\text{``}{\cong}\text{''}}}
\newcommand{\quasineq}{\mathrel{\text{``}{\neq}\text{''}}}

\newcommand{\cC}{\mathcal{C}}

\newcommand{\BBC}{\mathbb{C}}
\newcommand{\BBZ}{\mathbb{Z}}
\newcommand{\bq}{\boldsymbol{q}}
\newcommand{\bolds}{\boldsymbol{s}}

\newcommand{\bx}{\boldsymbol{x}}
\newcommand{\bby}{\boldsymbol{y}}
\newcommand{\bz}{\boldsymbol{z}}

\newcommand{\dd}{\partial}

\newcommand{\Id}{{\mathrm d}}

\begin{document}

\pagestyle{plain}

\title
{The right\/-\/hand side of the Jacobi identity:\\[2pt] 
\mbox{ }\qquad 
to be naught or not to be\,?}

\author{Arthemy V Kiselev}

\address{Johann Bernoulli Institute for Mathematics and Computer Science,
  University of Groningen,
  P.O.Box~407, 9700\,AK Groningen, The Netherlands}

\ead{A.V.Kiselev@rug.nl}

\begin{abstract}
The geometric approach~\cite{gvbv} to iterated variations of local functionals --\,e.g., of the (master-)\/action functional\,-- resulted in an extension of the deformation quantisation technique to the set\/-\/up of Poisson models of field theory~\cite{dq15}. It also allowed of a rigorous proof~\cite{gvbv,cycle14} for the main inter\/-\/relations between the Batalin\/--\/Vilkovisky (BV) Laplacian~$\Delta$ and variational Schouten bracket~$\lshad\,,\,\rshad$. The 
{ad hoc} use 
of these relations had been a long\/-\/standing analytic dif\-fi\-cul\-ty in the BV-\/formalism for quantisation of gauge systems; now achieved, the proof does actually not require the assumption of graded\/-\/commutativity~\cite{cycle14}.
Explained in our previous work~\cite{gvbv,dq15,cycle14}, geometry's self\/-\/regularisation is rendered 
by Gel'fand's calculus of singular linear integral operators supported on the diagonal.

We now illustrate that analytic technique by inspecting the validity mechanism~\cite{sqs13} for the graded Jacobi identity which the variational Schouten bracket does satisfy (whence $\Delta^2=0$, i.e., the BV-\/Laplacian is a differential acting in the algebra of local functionals). By using one tuple of three variational multi\/-\/vectors twice, we contrast the new logic of iterated variations --\,when the right\/-\/hand side of Jacobi's identity vanishes altogether\,-- with the old method: interlacing its steps and stops, it could produce some non\/-\/zero representative of the trivial class in the
top\/-\/degree 
horizontal cohomology. But we then show at once by an elementary counterexample why, in the frames of the old approach that did not rely on Gel'fand's calculus, the BV\/-\/Laplacian 
failed to be a graded derivation of the variational Schouten 
bracket
.\\[2pt]
\end{abstract}

\section{Introduction}
To describe Nature, we need analytic, geometric, and algebraic instruments.
Their formalism must simultaneously be mathematically rigorous and
yield relevant physical predictions.
One of such tools --\,to work with 
the geometry of iterated variations of 
local functionals in field theory models\,-- was designed 
in~\cite{gvbv,sqs13}, see also~\cite{dq15,cycle14}.
The aim of this paper is to illustrate by explicit examples of 
calculations, commented at every step, how this new technique is used 
in practice.
We refer to the previous report~\cite{gvbv} for basic theory outline in~\S1--2 therein, viewing it as a pre\/-\/requisite for 
the further exercise. Let us keep in mind that the calculus of singular linear integral operators~\cite{GelfandShilov} is the working language of that mathematical technique~--- itself serving the tasks of theoretical physics (cf.~\cite{GitmanTyutin,HenneauxTeitelboim}). That geometry's and formalism's noncommutative extension has been developed in~\cite[\S1--2]{cycle14} (see also~\cite{KontsevichCyclic} and~\cite{OlverSokolovCMP1998} in a similar context), whereas their implementation in the graded\/-\/commutative set\/-\/up is performed in~\cite[\S3]{gvbv} and~\cite[\S3]{dq15}. 
Besides the proof of 
theorems in~\cite{gvbv,cycle14,sqs13}, the reader will find there an overview of the history of the problem --\,e.g., in the context of Batalin\/--\/Vilkovisky's formalism for quantisation of gauge systems~\cite{BV1981
}\,-- accompanied with a list of literature references. The use of new technique in the lift of Kontsevich's deformation quantisation procedure~\cite{KontsevichFormality} to the geometry of field models is crucial in~\cite{dq15}. We finally recall that the master\/-\/functional realisation of zero\/-\/curvature representations for kinematically integrable systems was obtained in~\cite{Norway}, giving rise to the problem of BV-\/quantisation in the context of inverse scattering method. That line of research also relies on the technique which is exemplified in this note.


\subsection{The geometry of functionals: an overview}
Before motivating the choice of a class of examples to analyse, let us first recall an elementary still crucial distinction between the concepts of \emph{functions} on finite\/-\/dimensional smooth 
manifolds and, on the other hand, \emph{functionals} that appear in 
field theory models.\footnote{%
Not only the \emph{integral} functionals are then considered --\,e.g., such as the action $S=\int L\bigl(\bx,[\bq]\bigr)\cdot\dvol(\bx)\colon\Gamma(\pi)\to\Bbbk$ that takes the field configurations $\bigl\{\bq=\phi(\bx)$, $\bx\in M^m\bigr\}$ from the set $\Gamma(\pi)\ni\phi$ of sections for the (super)\/bundle~$\pi$ of a given physical model to the (complex) number field~$\Bbbk$.
The (sums of) formal products of integral functionals are also introduced, thus extending the vector space~$\bar{H}^m(\pi)\ni S$ to the algebra~$\overline{\mathfrak{M}}^m(\pi)$ of \emph{local} functionals.
For instance, the functional $\exp\bigl(\tfrac{\boldsymbol{i}}{\hbar}S([\bq])\bigr)=1+\tfrac{\boldsymbol{i}}{\hbar}S-\tfrac{1}{2\hbar^2}\,S\cdot S+\cdots\colon\Gamma(\pi)\to\BBC[[\hbar]]$, depending on the field configurations $\{\bq=\phi(\bx)\}$ and possibly, on the sections' derivatives (which is denoted by using the square brackets:~$[\ldots]$), serves as a weight factor in the construction of Feynman path integral.}
(In retrospect, the examples under study would have motivated the necessity to re\/-\/consider the geometry of iterated variations and find the proper analytic apparatus~\cite{gvbv} that did resolve the difficulties; the existence of conceptual difficulties in the 
formalism was of course known long ago, see~\cite[\S15.1]{HenneauxTeitelboim}.)

Specifically, a function $f\colon N^n\to\Bbbk$ on a finite\/-\/dimensional manifold~$N^n$ is the \emph{zero function} whenever its value at every point of the domain of definition is the zero~$0\in\Bbbk$. On the other hand, by integrating the physical field values~$\phi(\bx)\in N^n$ over points~$\bx\in M^m$ of the underlying space\/-\/time $\bigl(M^m$,\ $\dvol(\bx,\phi(\bx))\bigr)$, a given field model's functional $F\colon\Gamma(\pi)\to\Bbbk$ is a \emph{zero functional} if its value~$F\bigl([\phi]\bigr)$ is~$0\in\Bbbk$ for all configurations~$\phi\in\Gamma(\pi)$. Still is it readily seen that, provided that the topology of the bundle~$\pi$ and all the boundary conditions are properly taken into account (typically, either the boundary conditions are periodic or one assumes the sections' rapid decay towards the boundary~$\dd M^m$, if~any), a local functional would always attain the zero value at all sections~$\phi\in\Gamma(\pi)$ if at least one factor in every integral functionals' formal product is realised by an exact top\/-\/degree horizontal form~$\Id_h(\eta)$, so that $\int\Id_h(\eta)\cong\int0\in\bar{H}^m(\pi)$.
(Whenever the topology of~$\pi$ creeps in, one quotients out the respective invariants of the bundle~$\pi$; for instance, we have that $\int_{\mathbb{S}^1}\Id\alpha-2\pi\equiv0$ for all sections~$\phi\in\Gamma(\pi)$ of bundles over the circle~$M^1\mathrel{{:}{=}}\mathbb{S}^1$.)

However, we discovered in~\cite{gvbv} (cf.~\cite[\S3.4]{dq15}) that there is a crucial distinction between generic homologically\/-\/trivial integral factors~$\int\Id_h(\eta)\cong\int0\in\bar{H}^m(\pi)$ and the zero\/-\/density functional $\int0\cdot\dvol(\cdot,\phi(\cdot))$. This alternative 
echoes whenever a functional which manifestly yields zero for every section~$\phi\in\Gamma(\pi)$ appears at an intermediate but strictly not the last step of a reasoning.
Due to that distinction's mechanism, which will be re\/-\/addressed and illustrated in what follows, the \emph{synonyms for zero} $\int\Id_h(\eta)\cong\int0\in\bar{H}^m(\pi)$ can contribute nontrivially to 
reasoning's end\/-\/product. This paradoxical behaviour was explained within the geometric approach to iterated variations of local functionals~\cite{gvbv,cycle14}; the resolution did allow of a rigorous proof of formalism's main identities (e.g., compare~\eqref{EqZimes} on p.~\pageref{EqZimes}, which had been postulated ad hoc, with~\eqref{EqNotHolds} below).

The geometric mechanism behind that puzzling effect is based on the following fact. Each integral object in a given geometry of fibre (super)\/bundle~$\pi$ for 
fields~$\phi\in\Gamma(\pi)$ refers to its own copy of the base manifold~$M^m$ over which the integration by using the volume element~$\dvol(\cdot,\phi(\cdot))$ is performed. Represented by functionals $\Gamma(\pi)\to\Bbbk$, the theory's physical objects interact by serving as the input data for operators and structures such as the Batalin\/--\/Vilkovisky Laplacian~$\Delta$ or variational Schouten bracket~$\lshad\,,\,\rshad$; in this way, the new object is produced. In the frames of local field theory, the processing of input data entails the merging of all copies of the constintuent objects' integration domains~$M^m$, so that such functionals' densities --\,or their derivatives\,-- are then evaluated at the diagonal in the respective products~$M^m\times\ldots\times M^m$ of the base copies. (Moreover, the evaluation of all densities belonging to different input objects occurs at the graph of a given field portrait $\{\phi(\bx)$,\ $\bx\in M^m\}$, now shared by all the input objects' contributions.) When does the restriction to such diagonals take place\,? The answer is not obvious.
Whereas the full 
geometric picture 
keeps track of several copies $M^m\times\ldots\times M^m$
of the integration manifold~$M^m$ --\,in both
the functionals and 
(pairs of) field variations' parity\/-\/even and odd 
components,\,-- too early 
does the traditional approach merge the integration domains.
In effect, the results of two seemingly identical calculations can become unequal as colomology classes\footnote{%
For instance, the inequality $\lshad\cdot,\Delta(\cdot)\rshad(F,H)\ncong
\lshad\cdot,\cdot\rshad(F,\Delta H)$ \emph{can} be valid,
cf.\ conclusive Remark~\ref{RemSynonyms} on p.~\pageref{RemSynonyms}.}
---~for the same input data\,! 

Clearly, it is desirable that, serving the geometry of field theory, the calculus were consistent, not led to contradictions in the course of verification of formalism's main identities, and that it allowed of maningful predictions in the models at hand. The practical value of this note is that it helps 
to avoid inconsistencies or brough\/-\/in mismatches, hence preventing 
a necessity of postfactum regularisation by adding the various correction counterterms etc.

\subsection{Another look on the variation of functionals}
Now let us describe the class of identities which are examined and exemplified
in what follows. 
To this end, we recall that
the variational geometry of jet bundles $J^\infty\bigl(\pi\colon
E^{m+(n_0|n_1)}\xrightarrow{\ N^{(n_0|n_1)}\ }M^m\bigr)$
and the calculus of variational multi\/-\/vectors on jet spaces enlarge the symplectic geometry for usual \mbox{(super)} manifolds~$N^{(n_0|n_1)}$, 
parametrised locally by using parity\/-\/even and odd local coordinates~$(\bq,\bq^\dagger)$; one can integrate by parts in the new set\/-\/up. The classes of highest horizontal cohomology groups~$\bar{H}{}^m(\pi)$ are generated by the lift~$\Id_h$ of the de Rham differential on the base~$M^m$ to the total space~$J^\infty(\pi)$. Each $\Id_h$-\/cohomology class carries a substantial freedom: a $\Id_h$-\/exact term added to a class representative in the input data, nothing changes in the output of a calculation~--- provided that a variational derivative acts on that input term \emph{at once}.\footnote{Let us recall that the identity $\delta/\delta\bq\circ\Id/\Id x\equiv0$ is the primary exercise in the entire theory of variations.}

In these terms, the variational derivatives~$\delta/\delta\bq$ and~$\delta/\delta\bq^\dagger$ are viewed in the Lagrangian or Hamiltonian formalisms~\cite{KuperCotangent} as proper extensions of the partial derivatives~$\dd/\dd\bq$ and~$\dd/\dd\bq^\dagger$ along the fibre~$N^{(n_0|n_1)}$ in the (super-)\/bundle~$\pi$. These extensions are immediate indeed, meaning that all the integrations by parts over~$M^m$ are indivisibly attached to the derivations along the fibre.
The two analytic operations are not separated, even though the integrations could be postponed to a later moment. The resulting approach is adequate for the one\/-\/step derivation of Euler\/--\/Lagrange equations and for verification of the property for some systems of~PDEs to be manifestly Euler\/--\/Lagrange (by using the Helmholtz criterion, see~\cite{Olver}). The conventional formalism is sufficient also for the construction of variational Poisson brackets of integral functionals.\footnote{The variational Schouten bracket~$\lshad\,,\,\rshad$ itself is an example of variational Poisson bracket for the specific $\BBZ_2$-\/graded set\/-\/up such that the fibres~$N^{(n|n)}$ are described by using the parity\/-\/even coordinates~$\bq$ and their canonical conjugates~$\bq^\dagger$ of odd parity, cf.~\cite[\S1--2]{cycle14}.}

Strictly speaking, it follows from nowhere why the integrations by parts over the base manifold~$M^m$ should be inseparable from the partial derivatives along the fibres in~$J^\infty(\pi)$. It could well be that some other modus operandi is preferrable as soon as the transition from the supermanifold~$N^{(n|n)}$ standing alone to the super\/-\/bundle~$\pi$ with the fibre~$N^{(n|n)}$ over each point~$\bx\in M^m$ is accomplished.

An alternative approach has been reported in~\cite{gvbv}, see also~\cite[\S2.2]{cycle14} and~\cite[\S3.2]{dq15}. In brief, every calculation is split in two steps. First, all the partial 
derivatives $\dd/\dd\bq_\sigma$ and~$\dd/\dd\bq^\dagger_\tau$ act on the input functionals' densities. We remember that each input object brings its own copy of the base~$M^m$ into the emergent product of bundles (namely, there is one copy of the bundle~$\pi$ for each integral functional and there is a copy of the tangent bundle~$T\pi$ for each variation of~$\phi\in\Gamma(\pi)$ along the fibres of~$\pi$). This means that the newly constructed, intermediate objects still retain a kind of memory about the way how they have been produced. 
Now is the second step of a reasoning: the integration domains~$M^m$ in the products~$M^m\times\ldots\times M^m$ are merged by the restriction to the diagonal. The integrations by parts then produce the respective total derivatives $\pm\Id/\Id\bx^\sigma$ and~$\pm\Id/\Id\bx^\tau$ 
along the base~$M^m\ni\bx$ (e.g., contrast~\eqref{EqCombineTrue} on p.~\pageref{EqCombineTrue} with~\eqref{EqCombineWrong} on p.~\pageref{EqCombineWrong} below).


The Batalin\/--\/Vilkovisky technique for quantisation of gauge systems~\cite{BV1981
} puts the ma\-the\-ma\-ti\-cal methods of the calculus of variations to a hard test (cf.~\cite[\S3.2.3]{dq15}). The traditional approach~\cite{KuperCotangent} to variation of local functionals 
is incapable of avoiding contradictions in the calculus (e.g., see a counterexample on p.~\pageref{ExCounter} below; let us repeat that this class of difficulties had been known well in the literature~\cite[\S15.1]{HenneauxTeitelboim}). In con\-se\-quen\-ce, the claims in~\cite[\S1.3]{YKS2008} about 
interrelation between the Batalin\/--\/Vilkovisky Laplacian~$\Delta$ and variational Schouten bracket~$\lshad\,,\,\rshad$ are legitimate; they codify the picture that can be rigorously substantiated in the supergeometry of ordinary supermanifolds, but it is the jet\/-\/superbundle set\/-\/up in which there were neither proof nor examples (in fact, identity~\eqref{EqZimes} on p.~\pageref{EqZimes} below would have been the sumbling\/-\/stone).
The 
effect produced by our 
revision of action priorities is that the constructions of BV-\/formalism and the logic of its formulae resume working (in particular, justifying the claims made in~\cite{YKS2008}). 

The right\/-\/hand side of Jacobi's identity~\eqref{Jacobi4Schouten}
for the variational Schouten bracket~$\lshad\,,\,\rshad$ is a convenient indicator of the difference between the two approaches. By following the traditional scheme, one obtains some cohomologically trivial term in the right\/-\/hand side of the identity; that representative of the equivalence class of zero functional depends on a choice of re\-pre\-sen\-ta\-ti\-ves for the three inputs of the Jacobiator~$\Jac(\cdot,\cdot,\cdot)$ in the left\/-\/hand side. But within the geometric picture, the identity's right\/-\/hand side is the integral functional the density of which vanishes by construction. 
(It is the researcher who could then add to it at will any 
other trivial functional.)
The aim of this paper is to illustrate such 
distinction. 
For this, we also show that within the same (super)\/geometry 
yet in a class of elementary problems just next to the verification of Jacobi's identity, all rigour is destroyed if the old, step\/-\/by\/-\/step calculation recipe is used.
This is confirmed in this paper by using a counterexample 
(
the full matching of the same objects is demonstrated 
in~\cite{gvbv}
by using a ``counter\/-\/counterexample'').\\
\centerline{\rule{1in}{0.7pt}}

\medskip\noindent%
The structure of this paper parallels 
the 
length of formulae under consideration. Indeed,
the three functionals~$F$,\ $G$,\ and~$H$ which we use in all the examples in this text are such that the calculation of~$\Jac(F,G,H)=0$ within the true theory of variations is done by hand in~\S\ref{SecTrue}. The cohomology class estimate $\Jac\bigl(F(x),G(x),H(x)\bigr)\cong0$ in the frames of traditional approach would be also manageable by hand but hard 
(see Example~\ref{ExNaiveOneBase} in~\S\ref{SecNaive}), whereas the inspection of validity mechanism for Jacobi's identity via the restriction $\Jac\bigl(F(x),G(y),H(z)\bigr){\Bigr|}_{x=y=z}\cong0$ is fairly impossible without using proper software for symbolic computations~\cite{Jets,SsTools}, 
see Example~\ref{ExNaimeManyBases} on p.~\pageref{ExNaimeManyBases}. 
A short Counterexample~\ref{ExCounter} in~\S\ref{SecCounter}
takes us back to the main theorem in~\cite{gvbv} 
and its proper illustration (contained on pp.\,34--36 therein):
the full 
picture of iterated variations is a key to the resolution of apparent inconsistency. 

The notation is standard and as simple as possible. 
We shall work in the (graded-) commutative set\/-\/up; 
we refer to~\cite{cycle14} for its generalisation to a class of non\/-\/commutative geometries (see also references therein; the seminal paper is~\cite{KontsevichCyclic}).
In all examples, the base manifold~$M^m\ni\bx$ is one\/-\/dimensional and boundary terms are always discarded.\footnote{%
Whenever the model geometry is taken from the closed string or brane theory, the source manifold~$M^m$ is closed so that there is no boundary to have any such terms at (cf.~\cite{DubrZhang2001}). In the frames of field theory one postulates the rapid decay of every section~$\phi
\in\Gamma(\pi)$ at the space\/-\/time infinity~\cite{Olver,VinogradovCSpecII}, but it would take quite some effort to get there and bring back the minus sign from the integration by parts. However, the core idea 
of fields as excitations of the local degrees of freedom~\cite{GitmanTyutin,HenneauxTeitelboim,BV1981
} allows us to consider their test shifts~$\delta\bq\bigl(\bx,
\phi(\bx)\bigr)$ with compact supports concentrated only around a point of~$M^m$, whence improper integrals over the space\/-\/time make sense, reducing to proper integrals over such tiny neighbourhoods.}
The two fibre coordinates in the vector superbundle~$\pi$ are the parity\/-\/even~$q$ and its canonical conjugate, parity\/-\/odd~$q^\dagger$.
The volume form~$\dvol(x)$ in the integral functionals is just~$\Id x$ in the weak\/-\/field approximation.


\section{Geometric approach to the Jacobi identity}\label{SecTrue}
\subsection{The variational Schouten bracket}\label{SecSchouten}
Denote by~$\bx$ a local coordinate on the base~$M^m$ of the variational cotangent superbundle~$\pi$; denote by~$\bq$ and~$\bq^\dagger$ the $n$-\/tuples of respective parity\/-\/even and parity\/-\/odd canonical conjugate variables along the fibres of~$\pi$. Suppose $\bolds\in\Gamma(\pi)$ is a section. Denote by~$\vec{e}_i\bigl(\bx,\bolds(\bx)\bigr)$ 
and~$\vec{e}^{\,\dagger,i}\bigl(\bx,\bolds(\bx)\bigr)$ a smooth field of dual bases in the parity\/-\/even and odd halves of the spaces~$T_{(\bx,\bolds(\bx))}\pi^{-1}(\bx)$ tangent --\,at points~$(\bx,\bolds(\bx))$\,-- to the fibres over points~$\bx\in M^m$.
By construction, the two ordered couplings between elements of those bases are normalised: at every value of the index~$i$ we have that (no summation\,!)
\begin{equation}\label{EqCouplingsValues}
\bigl\langle 
\underrightarrow{
\stackrel{\text{first}}{\vec{e}_i\bigl(\bby_1,\bolds(\bby_1)\bigr)},
\stackrel{\text{second}}{\vec{e}^{\,\dagger,i}\bigl(\bby_2,\bolds(\bby_2)\bigr)}
}
\bigr\rangle {\Bigr|}_{\bby_1=\bby_2} = +1, \qquad
\bigl\langle 
\underrightarrow{
\stackrel{\text{first}}{\vec{e}^{\,\dagger,i}\bigl(\bby_2,\bolds(\bby_2)\bigr)},
\stackrel{\text{second}}{\vec{e}_i\bigl(\bby_1,\bolds(\bby_1)\bigr)}
}
\bigr\rangle {\Bigr|}_{\bby_1=\bby_2} = -1.
\end{equation}
For the sake of brevity, we shall not indicate 
the basic (co)\/vectors' dependence on points in the fibres of~$\pi$, writing only $\vec{e}_i(\bby)$ or~$\vec{e}^{\,\dagger,i}(\bz)$ from now on.

Consider an infinitesimal shift,
\[
\bigl(\delta\bq(\bby),\delta\bq^\dagger(\bby)\bigr)=
\sum\limits_{i=1}^n \Bigl(
\delta s^i\bigl(\bby,\bolds(\bby)\bigr)\cdot\vec{e}_i(\bby) +
\delta s^\dagger_i\bigl(\bby,\bolds(\bby)\bigr)\cdot\vec{e}^{\,\dagger,i}(\bby)
\Bigr),
\]
of the bipartite section $\bolds=\bigl(s^i(\bx),s^\dagger_i(\bx)\bigr)$ of the variational cotangent superbundle~$\pi$.
We postulate that for every~$i$ and over all points~$\bx\in M^m$ of the support
$\supp\delta s^i\bigl(\bx,\bolds(\bx)\bigr)$,
the respective components~$\delta s^i$ and~$\delta s^\dagger_i$ of the 
section $(\delta\bq,\delta\bq^\dagger)\in\Gamma(T\pi)$ are normalised by using the rule
\[
\delta s^i\bigl(\bx,\bolds(\bx)\bigr)\cdot\delta s^\dagger_i\bigl(\bx,\bolds(\bx)\bigr)\equiv 1\qquad\text{(no summation over~$i$)}.
\]
By convention, the vertical differentials (i.e., those involving only the derivations along the fibres of~$J^\infty(\pi)$) of all 
objects from the variational symplectic supergeometry are always expanded with respect to the elements~$+\vec{e}^{\,\dagger,i}(\cdot)$ and~$+\vec{e}_i(\cdot)$ of the mutually dual bases. Namely, such differentials are expressed by
\[
\Bigl(f\bigl(\bx_1,[\bq],[\bq^\dagger]\bigr)\Bigr)\,\smash{\frac{\overleftarrow{\dd}}{\dd q^i}}\cdot\vec{e}^{\,\dagger,i}(\bx_1)+\ldots \qquad\text{or}\qquad
\ldots+\vec{e}_i(\bx_2)\cdot\smash{\frac{\overrightarrow{\dd}}{\dd q^\dagger_i}}\,
\Bigl(g\bigl(\bx_2,[\bq],[\bq^\dagger]\bigr)\Bigr),
\]
where $f$ and~$g$ are densities of the integral objects to\/-\/vary.
Supported on the diagonal, the singular linear integral operators for integral objects' variation --\,in the course of infinitesimal shifts of the sections at which the objects will be evaluated\,-- are such that the couplings in~\eqref{EqCouplingsValues} always yield~$+1$. Specifically, we have that
\[
\overrightarrow{\delta\bolds}=
\int\Id\bby\,\Bigl{\langle}(\delta s^i)\Bigl(\frac{\overleftarrow{\dd}}{\dd\bby}\Bigr)^{\sigma}(\bby)\cdot
\underrightarrow{
\stackrel{\text{first}}{\vec{e}_i(\bby)}|\stackrel{\text{second}}{\vec{e}^{\mathstrut\,\dagger,i}(\,\cdot\,)}
}
\Bigr{\rangle}\,\frac{\overrightarrow{\dd}}{\dd q^i_{\sigma}},
\]
\begin{align*}
\overrightarrow{\delta\bolds^{\dagger}}=&
\int\Id\bby\,\Bigl{\langle}(\delta s^{\dagger}_i)\Bigl(\frac{\overleftarrow{\dd}}{\dd\bby}\Bigr)^{\sigma}(\bby)\cdot
\underrightarrow{
(\stackrel{\text{first}}{{-}\vec{e}^{\mathstrut\,\dagger,i})(\bby)}
|\stackrel{\text{second}}{\vec{e}_i(\,\cdot\,)}
}
\Bigr{\rangle}\,\frac{\overrightarrow{\dd}}{\dd q^\dagger_{i,\sigma}},
\\
\overleftarrow{\delta\bolds}=&
\int\Id\bby\,\frac{\overleftarrow{\dd}}{\dd q^i_{\sigma}}\,
\Bigl{\langle}
\underleftarrow{
\stackrel{\text{second}}{\vec{e}^{\mathstrut\,\dagger,i}(\,\cdot\,)}|\stackrel{\text{first}}{\vec{e}_i(\bby)}
}
\cdot
\Bigl(\frac{\overleftarrow{\dd}}{\dd\bby}\Bigr)^{\sigma}(\delta s^i)(\bby)
\Bigr{\rangle},
\\
\overleftarrow{\delta\bolds^{\dagger}}=&
\int\Id\bby\,\frac{\overleftarrow{\dd}}{\dd q^\dagger_{i,\sigma}}\,
\Bigl{\langle}
\underleftarrow{
\stackrel{\text{second}}{\vec{e}_i(\,\cdot\,)}|\stackrel{\text{first}}{(-\vec{e}^{\mathstrut\,\dagger,i})(\bby)}
}
\cdot
\Bigl(\frac{\overleftarrow{\dd}}{\dd\bby}\Bigr)^{\sigma}(\delta s^{\dagger}_i)(\bby)
\Bigr{\rangle},
\end{align*}
see~\cite[\S2.2--3]{gvbv} for details.\footnote{%
In particular, the matching of indexes~$i$ in the expansions of sections' shifts and objects' differentials reflects the fact that the bases~$\vec{e}_i$ and~$\vec{e}^{\,\dagger,i}$ are dual. Likewise, the matching of multi\/-\/indices~$\sigma$ refers to the definition of vector as equivalence class of trajectories passing through its attachment point.}

Let $F=\int f\bigl(\bx_1,[\bq],[\bq^\dagger]\bigr)\,\dvol(\bx_1)$ and
$G=\int g\bigl(\bx_2,[\bq],[\bq^\dagger]\bigr)\,\dvol(\bx_2)$ be two integral functionals in the variational symplectic supergeometry of jet bundle~$J^\infty(\pi)$ over~$\pi$. Consider their formal product~$F\cdot G$; let us analyse the construction of response $\bigl(\overrightarrow{\delta\bolds}\circ\overrightarrow{\delta\bolds^\dagger}\bigr)(F\cdot G)$ to 
consecutive variations of the parity\/-\/odd, then parity\/-\/even components of a section~$\bolds\in\Gamma(\pi)$. 
(Assume that the infinitesimal shifts $\delta s^i\bigl(\bby,\bolds(\bby)\bigr)$ and~$\delta s^\dagger_i\bigl(\bby,\bolds(\bby)\bigr)$ are given.)
By the graded Leibniz rule, we have that
\begin{align*}
\bigl(\overrightarrow{\delta\bolds}&\circ\overrightarrow{\delta\bolds^\dagger}\bigr)(F\cdot G)=
\overrightarrow{\delta\bolds}\Bigl(\overrightarrow{\delta\bolds^\dagger}(F)\cdot G + (-)^{|F|}F\cdot\overrightarrow{\delta\bolds^\dagger}(G)\Bigr)
={}\\
{}&=\bigl(\overrightarrow{\delta\bolds}\circ\overrightarrow{\delta\bolds^\dagger}\bigr)(F)\cdot G
+ (-)^{|F|}\overrightarrow{\delta\bolds}(F)\cdot\overrightarrow{\delta\bolds^\dagger}(G)
+ \overrightarrow{\delta\bolds^\dagger}(F)\cdot\overrightarrow{\delta\bolds}(G)
+ (-)^{|F|}F\cdot\bigl(\overrightarrow{\delta\bolds}\circ\overrightarrow{\delta\bolds^\dagger}\bigr)(G).
\end{align*}
Using a slightly counterintuitive lemma\footnote{%
Not only the parity\/-\/odd derivation~$\vec{\dd}$ overtakes a graded object~$F$ but also, whenever the derivation reaches its final location at the object's right side, the graded operator~$\vec{\dd}$ is \emph{destroyed} and replaced by the graded derivation acting in the right\/-\/to\/-\/left direction. Without such last step, the rule of signs would dictate the formula 
$\vec{\dd}\circ F=\vec{\dd}(F)+(-)^{|F|\cdot|\dd|}F\circ\vec{\dd}$.}
$\overrightarrow{\dd}/\dd\bq^\dagger(F)=(-)^{|F|-1}(F)\overleftarrow{\dd}/\dd\bq^\dagger$, let us reverse the direction in which the operators~$\overrightarrow{\delta\bolds}$ and~$\overrightarrow{\delta\bolds^\dagger}$ act on~$F$ in the second and third terms of the formula above; 
this yields\footnote{%
Further processing of the first and last terms in the formula at hand --\,that is, the on\/-\/the\/-\/diagonal reconfigurations of couplings and integrations by parts\,-- is 
completely analogous to the algorithm for dealing with the second and third terms, see below; the result is~\eqref{EqDeviationDerivation}.}
\[
{}
=\bigl(\overrightarrow{\delta\bolds}\circ\overrightarrow{\delta\bolds^\dagger}\bigr)(F)\cdot G
+ (-)^{|F|}\Bigl(
(F)\overleftarrow{\delta\bolds}\cdot\overrightarrow{\delta\bolds^\dagger}(G)
- (F)\overleftarrow{\delta\bolds^\dagger}\cdot\overrightarrow{\delta\bolds}(G)
\Bigr)
+ (-)^{|F|}F\cdot\bigl(\overrightarrow{\delta\bolds}\circ\overrightarrow{\delta\bolds^\dagger}\bigr)(G).
\]
Let us have a closer look on the difference of second and third terms: it is
\begin{align*}
\phantom{+}&\iint\Id\bby_1\dvol(\bx_1)\,\bigl(f(\bx_1,[\bq],\bq^\dagger])\bigr)
\frac{\overleftarrow{\dd}}{\dd q^{i_1}_{\sigma_1}}\,
\Bigl\langle\underleftarrow{\stackrel{\text{second}}{\vec{e}^{\,\dagger,i_1}(\bx_1)},\stackrel{\text{first}}{\vec{e}_{i_1}(\bby_1)}} \cdot
\Bigl(\frac{\overrightarrow{\dd}}{\dd\bby_1}\Bigr)^{\sigma_1}
(\delta s^{i_1})(\bby_1)\Bigr\rangle\cdot{}
\\
{}&{}\quad{}\cdot\iint\Id\bby_2
\Bigl\langle (\delta s^\dagger_{i_2})\Bigl(\frac{\overleftarrow{\dd}}{\dd\bby_2}\Bigr)^{\sigma_2}(\bby_2)\cdot
\underrightarrow{\stackrel{\text{first}}{(-\vec{e}^{\,\dagger,i_2})(\bby_2)},\stackrel{\text{second}}{\vec{e}_{i_2}(\bx_2)}}
\Bigr\rangle\,\frac{\overrightarrow{\dd}}{\dd q^\dagger_{i_2,\sigma_2}}\bigl(g(\bx_2,[\bq],[\bq^\dagger])\bigr)\,\dvol(\bx_2)-{}
\\
{}&-\iint\Id\bby_1\dvol(\bx_1)\,\bigl(f(\bx_1,[\bq],\bq^\dagger])\bigr)
\frac{\overleftarrow{\dd}}{\dd q^{\dagger}_{i_1,\sigma_1}}\,
\Bigl\langle\underleftarrow{\stackrel{\text{second}}{\vec{e}_{i_1}(\bx_1)},\stackrel{\text{first}}{(-\vec{e}^{\,\dagger,i_1})(\bby_1)}} \cdot
\Bigl(\frac{\overrightarrow{\dd}}{\dd\bby_1}\Bigr)^{\sigma_1}
(\delta s^\dagger_{i_1})(\bby_1)\Bigr\rangle\cdot{}
\\
{}&{}\quad{}\cdot\iint\Id\bby_2
\Bigl\langle (\delta s^{i_2})\Bigl(\frac{\overleftarrow{\dd}}{\dd\bby_2}\Bigr)^{\sigma_2}(\bby_2)\cdot
\underrightarrow{\stackrel{\text{first}}{\vec{e}_{i_2}(\bby_2)},\stackrel{\text{second}}{\vec{e}^{\,\dagger,i_2}(\bx_2)}}
\Bigr\rangle\,\frac{\overrightarrow{\dd}}{\dd q^{i_2}_{\sigma_2}}\bigl(g(\bx_2,[\bq],[\bq^\dagger])\bigr)\,\dvol(\bx_2).
\end{align*}
As it has been explained in~\cite{gvbv} and then in~\cite[\S2.5--6]{cycle14},
the conversion of two pairs of variations in $(F)\overleftarrow{\delta\bolds}\cdot\overrightarrow{\delta\bolds^\dagger}(G)
-(F)\overleftarrow{\delta\bolds^\dagger}\cdot\overrightarrow{\delta\bolds}(G)$
into one integral object --\,via integrations by parts on the diagonal $\bx_1=\bby_1=\bby_2=\bx_2$ through many consecutive reconfigurations of the couplings\,-- determines the 
functional\footnote{%
The remaining volume element can be either~$\dvol(\bx_1)$ or~$\dvol(\bx_2)$; 
its final location is prescribed by either the right\/-\/to\/-\/left or left\/-\/to\/-\/right (which is the case here) direction of couplings in the output.
From~\eqref{EqCouplingsValues} it is clear that a simultaneous swap ``first~$\rightleftarrows$ second'' in a pair of couplings would give the extra factor $(-1)\cdot(-1)=+1$, so that expression's overall sign does not change.}
\begin{align}
\phantom{+}&\iiiint\Id\bx_1\Id\bby_1\Id\bby_2\,
\bigl(f(\bx_1,[\bq],\bq^\dagger])\bigr)
\frac{\overleftarrow{\dd}}{\dd q^{i_1}_{\sigma_1}}
\vphantom{\Bigl{|}}^{\lceil}
\Bigl(-\frac{\overleftarrow{\Id}}{\Id\bby_1}\Bigr)^{\sigma_1}
\vphantom{\Bigr{|}}^{\rceil}
\,\Bigl\langle\underline{\stackrel{\text{first}}{\vec{e}^{\,\dagger,i_1}(\bx_1)}}\Bigr|\cdot{}\notag
\\
&{}\qquad\qquad{}\cdot
\Bigl\langle \delta s^{i_1}(\bby_1)\cdot
\underline{\underline{\stackrel{\text{first}}{\vec{e}_{i_1}(\bby_1)}}},
\underline{\underline{\stackrel{\text{second}}{(-\vec{e}^{\,\dagger,i_2})(\bby_2)}}} \cdot \delta s^\dagger_{i_2}(\bby_2) \Bigr\rangle \cdot{}\notag
\\
&{}\qquad\qquad\qquad{}\cdot
\Bigl|\underline{\stackrel{\text{second}}{\vec{e}_{i_2}(\bx_2)}}
\Bigr\rangle\cdot{}
\vphantom{\Bigl{|}}^{\lceil}
\Bigl(-\frac{\overrightarrow{\Id}}{\Id\bby_2}\Bigr)^{\sigma_2}
\vphantom{\Bigr{|}}^{\rceil}
\frac{\overrightarrow{\dd}}{\dd q^\dagger_{i_2,\sigma_2}}\bigl(g(\bx_2,[\bq],[\bq^\dagger])\bigr)\,\dvol(\bx_2)-{}\notag
\\
{}&-
\iiiint\Id\bx_1\Id\bby_1\Id\bby_2\,
\bigl(f(\bx_1,[\bq],\bq^\dagger])\bigr)
\frac{\overleftarrow{\dd}}{\dd q^{\dagger}_{i_1,\sigma_1}}
\vphantom{\Bigl{|}}^{\lceil}
\Bigl(\frac{\overleftarrow{\Id}}{\Id\bby_1}\Bigr)^{\sigma_1}
\vphantom{\Bigr{|}}^{\rceil}
\,\Bigl\langle\underline{\stackrel{\text{first}}{\vec{e}_{i_1}(\bx_1)}}\Bigr|\cdot{}\notag
\\
&{}\qquad\qquad{}\cdot
\Bigl\langle 
\delta s^\dagger_{i_1}(\bby_1)
\underline{\underline{\stackrel{\text{first}}{(-\vec{e}^{\,\dagger,i_1})(\bby_1)}}},
\underline{\underline{\stackrel{\text{second}}{\vec{e}_{i_2}(\bby_2)}}}
\cdot \delta s^{i_2}(\bby_2)\Bigr\rangle \cdot{}\notag
\\
&{}\qquad\qquad\qquad{}\cdot
\Bigl|
\underline{\stackrel{\text{second}}{\vec{e}^{\,\dagger,i_2}(\bx_2)}}
\Bigr\rangle\,\cdot
\vphantom{\Bigl{|}}^{\lceil}
\Bigl(-\frac{\overrightarrow{\Id}}{\Id\bby_2}\Bigr)^{\sigma_2}
\vphantom{\Bigr{|}}^{\rceil}
\frac{\overrightarrow{\dd}}{\dd q^{i_2}_{\sigma_2}}\bigl(g(\bx_2,[\bq],[\bq^\dagger])\bigr)\,\dvol(\bx_2).\label{EqUniqueDef}
\end{align}
Evaluating both couplings in the minuend, we obtain $(-1)\cdot(-1)=+1$;
likewise, in the subtrahend we have that $(+1)\cdot(+1)=+1$; at every value of the indexes, the respective shift components contribute with $\delta s^\bullet\cdot\delta s^\dagger_\bullet=1$.
In effect, the only minus sign making the \emph{difference} of two terms is determined by the precedence $\bq\prec\bq^\dagger$ versus succedence $\bq^\dagger\succ\bq$, that is, by the sequential order in which the parity\/-\/even and odd partial derivatives along the fibres of~$J^\infty(\pi)$ are distributed between the arguments $F\prec G$, which were the ordered pair of input objects.
It is now readily seen that the operational algorithm for reconfiguration of normalised couplings in $\bigl(\overrightarrow{\delta\bolds}\circ\overrightarrow{\delta\bolds^\dagger}\bigr)(F\cdot G)$, which we started with, establishes the definition of variational Schouten bracket~$\lshad\,,\,\rshad$
as the descendent structure to the Batalin\/--\/Vilkovisky Laplacian~$\Delta$, cf.~\eqref{EqBVNaive} on p.~\pageref{EqBVNaive}:
\begin{equation}\label{EqDeviationDerivation}
\Delta(F\cdot G) \stackrel{\text{def}}{{}={}} 
\Delta F\cdot G + (-)^{|F|}\lshad F,G\rshad +(-)^{|F|}F\cdot\Delta G.
\end{equation}
In effect, the variational Schouten bracket~$\lshad\,,\,\rshad$ 
measures the de\-vi\-a\-ti\-on of~$\Delta$ from being a derivation.
Simultaneously, the variational Schouten bracket~$\lshad\,,\,\rshad$ extends the commutator~$[\,,\,]$ of (variational) one\/-\/vectors to the space of variational multi\/-\/vectors.\footnote{\label{FootCommutator}
The sequential order in which the densities of two arguments in~$\lshad\,,\,\rshad$ are differentiated with respect to the parity\/-\/even jet coordinates~$q^i_\sigma$ and parity\/-\/odd variables~$q^\dagger_{j,\tau}$ is often chosen in such a way that the shifted\/-\/graded skew\/-\/symmetric Schouten bracket of variational one\/-\/vectors $F=\int X^i\bigl(\bx,[\bq]\bigr)\,q^\dagger_i\,\dvol(\bx)$ and $G=\int Y^j\bigl(\bby,[\bq]\bigr)\,q^\dagger_j\,\dvol(\bby)$ is determined by \emph{minus} the usual commutator of the respective evolutionary vector fields within the purely even geometry of variables~$\bq$: one has that $\lshad F,G\rshad = - \int [X,Y]^k \bigl(\bz,[\bq]\bigr)\, q^\dagger_k\,\dvol(\bz)$, where $[X,Y]^k = + \bigl(X(Y^k)-Y(X^k)\bigr)$ is the componentwise action; this convention is adopted in~\cite{gvbv,sqs13}. The two conventions for~$[\,,\,]$ and~$\lshad\,,\,\rshad$ coincide if one takes $[X,Y]^k = - \bigl(X(Y^k)-Y(X^k)\bigr)$ as in Lie theory.}

However, we emphasize that the expression~$\lshad F,G\rshad$, which has been constructed by following the couplings' re\/-\/attachment mechanism, itself can serve as a constituent part of a larger object. Because the reconfigurations of couplings and integrations by parts occur prior only to the restriction of output to the jet of a section~$\bolds\in\Gamma(\pi)$, 
this would mean that the horizontal derivatives along the base~$M^m$ have not yet been brought to their anticipated final locations.
In the meantime, other partial derivatives
can freely overtake them.\footnote{We refer 
to~\cite{sqs13} for a demonstration how these conventions work in the proof of Jacobi's identity for the variational Schouten bracket~$\lshad\,,\,\rshad$.}
This is precisely why the total derivatives were embraced by using 
$\lceil\ldots\rceil$ and why the shifts' own base variables~$\bby_i$ were used instead of~$\bx_i$ from the functionals.
The delayed integration by parts guarantees that the iterated variations are \emph{graded\/-\/permutable}, stemming from the terms like this:
\begin{equation}\label{EqCombineTrue}
\Bigl(-\frac{\overrightarrow{\Id}}{\Id\bx}\Bigr)^{\sigma\cup\tau}\circ
\frac{\overrightarrow{\dd^2}}{\dd\bq_\tau\dd\bq^\dagger_\sigma},
\end{equation}
or similar~--- with any other combination of variables~$\bq$ and~$\bq^\dagger$.

Suppose still that the bracket of functionals~$F$ and~$G$ is the endpoint of a calculation (that is, the reasoning stops there and the object~$\lshad F,G \rshad\colon\Gamma(\pi)\to\Bbbk$ is used only for its evaluation at sections but it is \emph{not} contained 
in any larger formula such as the left\/-\/hand side of Jacobi's identity for~$\lshad\,,\,\rshad$). Should this be known in advance, then one re\/-\/derives the familiar pro\-vi\-si\-o\-nal formula (in fact, one of many~-- see~\cite{RingersProtaras}),
\begin{equation}\label{EqFamiliar}
\lshad F,G\rshad \quasieq
\int\Bigl( (f)\tfrac{\overleftarrow{\delta}}{\delta\bq}\cdot
\tfrac{\overrightarrow{\delta}}{\delta\bq^\dagger} (g) 
- 
(f)\tfrac{\overleftarrow{\delta}}{\delta\bq^\dagger}\cdot
\tfrac{\overrightarrow{\delta}}{\delta\bq} (g) \Bigr)\,\dvol(\bx).
\end{equation}
We recall that a step\/-\/by\/-\/step construction of objects which are then evaluated at sections is typical in the search for stationary points of action functionals in the Lagrangian formalism. This may not be the case in a larger framework.

\subsection{The Jacobi identity for~$\lshad\,,\,\rshad$}
The Jacobi identity for the Schouten bracket~$\lshad\,,\,\rshad$ can be understood as the graded Leibniz rule
\begin{equation}\label{Jacobi4Schouten}
\lshad F, \lshad G,H\rshad \rshad = \lshad \lshad F,G\rshad,H\rshad +
(-)^{(|F|-1)(|G|-1)}\,\lshad G,\lshad F,H\rshad\rshad;
\end{equation}
the bracket's own grading equals~$-1$, which is responsible for the shifts of grading~$|\cdot|$ in the exponent. Equivalently, the Jacobi identity for~$\lshad\,,\,\rshad$ is the (shifted-)\/graded commutator of operators~$\lshad F,\cdot\rshad$ and~$\lshad G,\cdot\rshad$ acting on a test functional~$H$,
\[
\lshad F,\lshad G,{\cdot}\rshad\rshad(H) - (-)^{(|F|-1)(|G|-1)}\,
\lshad G,\lshad F,{\cdot}\rshad\rshad(H) = 
\lshad\lshad F,G\rshad,{\cdot}\rshad(H);
\]
we refer to~\cite[\S2.6]{cycle14} and~\cite{sqs13} for the proof and many essential details.
\footnote{The bi\/-\/linear, shifted\/-\/graded skew\/-\/symmetric
structure~$\lshad\,,\,\rshad$ extends via
\[
\lshad F,G\cdot H\rshad = \lshad F,G\rshad\cdot H 
+ (-)^{(|F|-1)\cdot|G|} G\cdot\lshad F,H\rshad
\]
to the vector space of formal products $H_1\cdot\ldots\cdot H_\ell\colon\Gamma(\pi)\to\Bbbk$ of integral functionals.}


\begin{example}
Let us illustrate the validity mechanism for Jacobi's identity~\eqref{Jacobi4Schouten} by verifying it at three given
functionals. For simplicity, let there be just one independent variable~$x$, 
one parity\/-\/even coordinate~$q$ and its 
parity\/-\/odd canonical conjugate~$q^{\dagger}$. Set 
\[
F=\int q^{\dagger}qq_{x_1x_1}\,\dvol(x_1),\quad
G=\int q^{\dagger}_{x_2}\exp(q_{x_2})\,\dvol(x_2),\quad
\text{and } H=\int q^{\dagger}_{x_3x_3}\cos q\,\dvol(x_3);
\]
we note that the functionals $F$ and~$H$ re\/-\/appear in
Counterexample~\ref{ExCounter} on p.~\pageref{ExCounter} below and in the resolution to the paradox contained in it, see~\cite[pp.\,34--36]{gvbv}. 
We have $|F|=1$ and $|G|=1$, whence 
$(-)^{(|F|-1)(|G|-1)}=+1$ in~\eqref{Jacobi4Schouten}.
 
Let $\delta\bolds_1=(\delta s_1,\delta s_1^{\dagger})$ and 
$\delta\bolds_2=(\delta s_2,\delta s_2^{\dagger})$ be two normalized test shifts, i.\,e., suppose that 
$\delta s_{\alpha}(y)\cdot\delta s_{\alpha}^{\dagger}(y)=1$
at every~$y$ for $\alpha=1,2$. We recall from Lemma~1 in~\cite[p.~24]{gvbv} 
that the values of Schouten brackets
in~\eqref{Jacobi4Schouten} are independent of a concrete choice of the normalized functional coefficients 
$\delta s_{\alpha}$ and $\delta s^{\dagger}_{\alpha}$, which implies that the test shifts $\delta\bolds_1$ and $\delta\bolds_2$
in the inner and outer brackets can be swapped (this would amount to relabelling $y\rightleftarrows z$ of their arguments).
We shall 
not write the basic (co)\/vectors $\vec{e}(y)$ and
$\vec{e}^{\mathstrut\,\,\dagger}(y)$ in expansions of the test shifts and of the differentials of functionals' densities; it is enough to know the 
contributing values, which are specified by the precedence~$\bq\prec\bq^\dagger$ or antecedence~$\bq^\dagger\succ\bq$ and which are equal to~$\pm1$ 
(see \S\ref{SecSchouten} above
).

We have that%
\footnote{Let us repeat 
that integrations by parts, which cast the derivatives off the test shifts, are performed only 
when all the objects --\,such as the l.-h.s.\ or r.-h.s.\ of~\eqref{Jacobi4Schouten}\,-- are fully composed, all partial
derivatives of the functionals' densities are calculated, and reconfigurations of the couplings are ready to start.
In practice, this means that partial derivatives like $\overrightarrow{\dd}\!\!/\dd q_x$ or 
$\overleftarrow{\dd}\!\!/\dd q_{xx}^{\dagger}$ dive under $\overrightarrow{\Id}\!\!/\Id y$ or 
$\overleftarrow{\Id}\!\!/\Id z$ 
because those total derivatives 
have not yet appeared at the places where we write them ahead of time.}
$\lshad G,H\rshad={}$
\begin{multline*}
\smash{\iiiint} \Id y_2\,\Id y_3\,\Id x_2\,\dvol(x_3)\cdot
\Bigl\{\Bigl\langle
\vphantom{\bigl{|}}^{\lceil}
-\tfrac{\Id}{\Id y_2}
\vphantom{\bigr{|}}^{\rceil}
\bigl(\underbrace{ q^\dagger_{x_2}\,\exp(q_{x_2}) }_{x_2}\bigr)\cdot
\vphantom{\bigl{|}}^{\lceil}
\tfrac{\Id^2}{\Id y_3^2}
\vphantom{\bigr{|}}^{\rceil}
\,(\underbrace{ \cos q }_{x_3})\Bigr\rangle \cdot
\underbrace{\langle \delta s(y_2),\delta s^\dagger(y_3) \rangle}_{+1}
+{}\\
+\Bigl\langle \vphantom{\bigl{|}}^{\lceil}
-\tfrac{\Id}{\Id y_2}
\vphantom{\bigr{|}}^{\rceil}
{ \bigl(\underbrace{\exp(q_{x_2})}_{x_2}\bigr)\cdot\underbrace{q^\dagger_{x_3x_3}\cdot(-\sin q)}_{x_3} }\Bigr\rangle
\cdot \smash{ \underbrace{\langle \delta s^\dagger(y_2),\delta s(y_3) \rangle}_{-1} }
\Bigr\};
\end{multline*}
the integration variables~$x_i$ are displayed under the remnants of respective densities. Next, we obtain that
$\lshad F,\lshad G,H\rshad\rshad={}$
\begin{align*}
{}&\smash{ \int\!\Id z_1\int\!\Id z_{23}
\int\!\Id y_2\int\!\Id y_3\int\!\Id x_1\int\!\Id x_2\int\!\dvol(x_3) }\cdot
\underbrace{\langle \delta s(z_1),\delta s^\dagger(z_{23})\rangle}_{+1}\cdot{}
\\
{}&\ \Bigl\{
\Bigl\langle\bigl(\underbrace{{}^{\langle 1\rangle}\ %
q^\dagger q_{x_1x_1} + {}^{\langle 2\rangle}\ %
\tfrac{\Id^2}{\Id z_1^2}(q^\dagger q)}_{x_1}\bigr) \cdot
\bigl(-\tfrac{\Id}{\Id z_{\boldsymbol{2}3}}\bigr)
\bigl(-\tfrac{\Id}{\Id y_2}\bigr)\bigl(\underbrace{\exp(q_{x_2})}_{x_2}\bigr)
\cdot \tfrac{\Id^2}{\Id y_3^2} (\underbrace{\cos q}_{x_3})\Bigr\rangle \cdot
\underbrace{\langle \delta s(y_2),\delta s^\dagger(y_3)\rangle}_{+1} +{}
\\
{}&{}\quad
+\Bigl\langle\bigl(\underbrace{{}^{\langle 3\rangle}\ %
q^\dagger q_{x_1x_1} + {}^{\langle 4\rangle}\ %
\tfrac{\Id^2}{\Id z_1^2}(q^\dagger q)}_{x_1}\bigr) \cdot
\bigl(-\tfrac{\Id}{\Id y_2}\bigr)\bigl(\underbrace{\exp(q_{x_2})}_{x_2}\bigr)
\cdot 
\tfrac{\Id^2}{\Id z_{2\boldsymbol{3}}^2 } 
(\underbrace{-\sin q}_{x_3})\Bigr\rangle \cdot
\underbrace{\langle \delta s^\dagger(y_2),\delta s(y_3)\rangle}_{-1} 
\Bigr\} +{}
\\
{}&{}+\int\!\Id z_1\int\!\Id z_{23}
\int\!\Id y_2\int\!\Id y_3\int\!\Id x_1\int\!\Id x_2\int\!\dvol(x_3)\cdot
\underbrace{\langle \delta s^\dagger(z_1),\delta s(z_{23})\rangle}_{-1}\cdot{}
\\
{}&\ \Bigl\{\Bigl\langle
\bigl(\underbrace{{}^{\langle 5\rangle}\ q q_{x_1x_1} }_{x_1}\bigr) \cdot
\bigl(-\tfrac{\Id}{\Id z_{\boldsymbol{2}3}}\bigr)
\bigl(-\tfrac{\Id}{\Id y_2}\bigr)\bigl(\underbrace{q^\dagger_{x_2}\,\exp(q_{x_2})}_{x_2}\bigr)
\cdot \tfrac{\Id^2}{\Id y_3^2} (\underbrace{\cos q}_{x_3}) +{}
\\
{}&{}\qquad\qquad 
+ \bigl(\underbrace{{}^{\langle 6\rangle}\ q q_{x_1x_1} }_{x_1}\bigr) \cdot
\bigl(-\tfrac{\Id}{\Id y_2}\bigr)\bigl(\underbrace{q^\dagger_{x_2}\,\exp(q_{x_2})}_{x_2}\bigr)
\cdot 
\tfrac{\Id^2}{\Id y_3^2 } 
(\underbrace{-\sin q}_{x_3})
\Bigr\rangle\cdot
\underbrace{\langle \delta s(y_2),\delta s^\dagger(y_3)\rangle}_{+1} +{}
\\
{}&{}\quad
+\Bigl\langle
\bigl(\underbrace{{}^{\langle 7\rangle}\ q q_{x_1x_1} }_{x_1}\bigr) \cdot
\bigl(-\tfrac{\Id}{\Id z_{\boldsymbol{2}3} }\bigr)
\bigl(-\tfrac{\Id}{\Id y_2}\bigr)\bigl(\underbrace{\exp(q_{x_2})}_{x_2}\bigr)
\cdot (\underbrace{ q^\dagger_{x_3x_3}\cdot (-\sin q)}_{x_3}) +{}
\\
{}&{}\qquad\qquad 
{}+\bigl(\underbrace{{}^{\langle 8\rangle}\ q q_{x_1x_1} }_{x_1}\bigr) \cdot
\bigl(-\tfrac{\Id}{\Id y_2}\bigr) \bigl(\underbrace{\exp(q_{x_2})}_{x_2}\bigr)
\cdot (\underbrace{ q^\dagger_{x_3x_3}\cdot (-\cos q)}_{x_3})
\Bigr\rangle\cdot
\underbrace{\langle \delta s^\dagger(y_2),\delta s(y_3)\rangle}_{-1} \Bigr\}.
\end{align*}
On the other hand, $\lshad F,G\rshad={}$
\begin{multline*}
\iiiint \Id y_1\,\Id y_2\,\Id x_1\,\dvol(x_2)\cdot
\Bigl\{
\Bigl\langle
\bigl(\underbrace{q^\dagger q_{x_1x_1} + 
\tfrac{\Id^2}{\Id y_1^2}(q^\dagger q)}_{x_1}\bigr) \cdot
\vphantom{\bigl{|}}^{\lceil}
-\tfrac{\Id}{\Id y_2}
\vphantom{\bigr{|}}^{\rceil}
\bigl(\underbrace{\exp(q_{x_2})}_{x_2}\bigr)
\Bigr\rangle \cdot
\underbrace{\langle \delta s(y_1),\delta s^\dagger(y_2) \rangle}_{+1}
+{}\\
+\Bigl\langle \bigl(\underbrace{ q q_{x_1x_1} }_{x_1}\bigr) \cdot
\vphantom{\bigl{|}}^{\lceil}
-\tfrac{\Id}{\Id y_2}
\vphantom{\bigr{|}}^{\rceil}
\bigl(\underbrace{q^\dagger_{x_2}\,\exp(q_{x_2})}_{x_2}\bigr) \Bigr\rangle \cdot 
\underbrace{\langle \delta s^\dagger(y_1),\delta s(y_2) \rangle}_{-1}
\Bigr\}.
\end{multline*}
We infer that $\lshad\lshad F,G\rshad,H\rshad={}$
\begin{align*}
{}&\int\!\Id z_{12}\int\!\Id z_3
\int\!\Id y_1\int\!\Id y_2\int\!\Id x_1\int\!\Id x_2\int\!\dvol(x_3)\cdot
\underbrace{\langle \delta s(z_{12}),\delta s^\dagger(z_3)\rangle}_{+1}\cdot{}
\\
{}&\ \Bigl\{
\Bigl\langle \bigl( \underbrace{ {}^{\langle 9\rangle}\ %
\tfrac{\Id^2}{\Id z^2_{\boldsymbol{1}2}} (q^\dagger)+ 
{}^{\langle 10\rangle}\ %
\tfrac{\Id^2}{\Id y_1^2} (q^\dagger) }_{x_1} \bigr) \cdot
\bigl(-\tfrac{\Id}{\Id y_2}\bigr) \bigl(\underbrace{\exp(q_{x_2})}_{x_2}\bigr)
\cdot \tfrac{\Id^2}{\Id z_3^2 } (\underbrace{\cos q}_{x_3}) +{}
\\
{}&{}\qquad\quad 
+ \bigl(\underbrace{{}^{\langle 1\rangle}\ %
q^\dagger q_{x_1x_1} + {}^{\langle 2\rangle}\ %
\tfrac{\Id^2}{\Id y_1^2}(q^\dagger q)}_{x_1}\bigr) \cdot
\bigl(-\tfrac{\Id}{\Id z_{1\boldsymbol{2}}}\bigr)
\bigl(-\tfrac{\Id}{\Id y_2}\bigr)\bigl(\underbrace{\exp(q_{x_2})}_{x_2}\bigr)
\cdot \tfrac{\Id^2}{\Id z_3^2} (\underbrace{\cos q}_{x_3})
\Bigr\rangle \cdot
\underbrace{\langle \delta s(y_1),\delta s^\dagger(y_2)\rangle}_{+1} +{}
\end{align*}
\begin{align*}
{}&{}\quad
+\Bigl\langle
\bigl(\underbrace{{}^{\langle 11\rangle}\ %
q_{x_1x_1} + {}^{\langle 12\rangle}\ %
\tfrac{\Id^2}{\Id z_{\boldsymbol{1}2}^2}(q)}_{x_1}\bigr) \cdot
\bigl(-\tfrac{\Id}{\Id y_2}\bigr)\bigl(\underbrace{q^\dagger_{x_2}\,\exp(q_{x_2})}_{x_2}\bigr)
\cdot \tfrac{\Id^2}{\Id z_3^2} (\underbrace{\cos q}_{x_3}) + {}
\\
{}&{}\qquad\quad 
+ \bigl(\underbrace{{}^{\langle 5\rangle}\ q q_{x_1x_1} }_{x_1}\bigr) \cdot
\bigl(-\tfrac{\Id}{\Id z_{1\boldsymbol{2}}}\bigr)
\bigl(-\tfrac{\Id}{\Id y_2}\bigr)\bigl(\underbrace{q^\dagger_{x_2}\,\exp(q_{x_2})}_{x_2}\bigr)
\cdot \tfrac{\Id^2}{\Id z_3^2} (\underbrace{\cos q}_{x_3})
\Bigr\rangle \cdot
\underbrace{\langle \delta s^\dagger(y_1),\delta s(y_2)\rangle}_{-1}
\Bigr\} + {}
\\
{}&{}+\int\!\Id z_{12}\int\!\Id z_3
\int\!\Id y_1\int\!\Id y_2\int\!\Id x_1\int\!\Id x_2\int\!\dvol(x_3)\cdot
\underbrace{\langle \delta s^\dagger(z_{12}),\delta s(z_3)\rangle}_{-1}\cdot{}
\\
{}&\ \Bigl\{
\Bigl\langle
\bigl(\underbrace{{}^{\langle 13\rangle}\ %
q_{x_1x_1} + {}^{\langle 14\rangle}\ %
\tfrac{\Id^2}{\Id y_{1}^2}(q)}_{x_1}\bigr) \cdot
\bigl(-\tfrac{\Id}{\Id y_2}\bigr)\bigl(\underbrace{\exp(q_{x_2})}_{x_2}\bigr)
\cdot (\underbrace{ q^\dagger_{x_3x_3}\cdot (-\sin q)}_{x_3})
\Bigr\rangle \cdot
\underbrace{\langle \delta s(y_1),\delta s^\dagger(y_2)\rangle}_{+1} +{}
\\
{}&{}\quad
+\Bigl\langle
\bigl(\underbrace{{}^{\langle 7\rangle}\ q q_{x_1x_1} }_{x_1}\bigr) \cdot
\bigl(-\tfrac{\Id}{\Id z_{1\boldsymbol{2}} }\bigr)
\bigl(-\tfrac{\Id}{\Id y_2}\bigr)\bigl(\underbrace{\exp(q_{x_2})}_{x_2}\bigr)
\cdot (\underbrace{ q^\dagger_{x_3x_3}\cdot (-\sin q)}_{x_3})
\Bigr\rangle\cdot
\underbrace{\langle \delta s^\dagger(y_1),\delta s(y_2)\rangle}_{-1}
\Bigr\}.
\end{align*}
Thirdly, $\lshad F,H\rshad={}$
\begin{multline*}
\iiiint \Id y_1\,\Id y_3\,\Id x_1\,\dvol(x_3)\cdot
\Bigl\{
\Bigl\langle
\bigl(\underbrace{ q^\dagger q_{x_1x_1} + 
\tfrac{\Id^2}{\Id y_1^2}(q^\dagger q)}_{x_1}\bigr) \cdot
\vphantom{\bigl{|}}^{\lceil}
\tfrac{\Id^2}{\Id y_3^2}
\vphantom{\bigr{|}}^{\rceil}
(\underbrace{\cos q}_{x_3})
\Bigr\rangle \cdot
\underbrace{\langle \delta s(y_1),\delta s^\dagger(y_3) \rangle}_{+1}
+{}\\
+\Bigl\langle 
\bigl(\underbrace{ q q_{x_1x_1} }_{x_1}\bigr) \cdot
(\underbrace{ q^\dagger_{x_3x_3}\cdot (-\sin q)}_{x_3})
\Bigr\rangle \cdot 
\underbrace{\langle \delta s^\dagger(y_1),\delta s(y_3) \rangle}_{-1}
\Bigr\}.
\end{multline*}
In view of the functionals' gradings, we have
$+1\cdot\lshad G,\lshad F,H\rshad\rshad={}$
\begin{align*}
{}&\int\!\Id z_2\int\!\Id z_{13}
\int\!\Id y_1\int\!\Id y_3\int\!\Id x_1\int\!\Id x_2\int\!\dvol(x_3)\cdot
\underbrace{\langle \delta s(z_2),\delta s^\dagger(z_{13})\rangle}_{+1}\cdot{}
\\
{}&\ \Bigl\{\Bigl\langle
\bigl(-\tfrac{\Id}{\Id z_2}\bigr)\bigl(\underbrace{q^\dagger_{x_2}\,\exp(q_{x_2})}_{x_2}\bigr)
\cdot \bigl(\underbrace{{}^{\langle 11\rangle}\ %
q_{x_1x_1} + {}^{\langle 12\rangle}\ %
\tfrac{\Id^2}{\Id y_1^2}(q)}_{x_1}\bigr) \cdot
\tfrac{\Id^2}{\Id y_3^2} (\underbrace{\cos q}_{x_3})
\Bigr\rangle \cdot
\underbrace{\langle \delta s(y_1),\delta s^\dagger(y_3) \rangle}_{+1} +{}
\\
{}&{}\qquad\quad 
+\Bigl\langle
\bigl(-\tfrac{\Id}{\Id z_2}\bigr)\bigl(\underbrace{q^\dagger_{x_2}\,\exp(q_{x_2})}_{x_2}\bigr) \cdot 
\bigl(\underbrace{{}^{\langle 6\rangle}\ q q_{x_1x_1} }_{x_1}\bigr) \cdot
\tfrac{\Id^2}{\Id z_{1\boldsymbol{3}}^2 } (\underbrace{-\sin q}_{x_3})
\Bigr\rangle \cdot 
\underbrace{\langle \delta s^\dagger(y_1),\delta s(y_3) \rangle}_{-1}
\Bigr\} +{}
\\
{}&{}+\int\!\Id z_2\int\!\Id z_{13}
\int\!\Id y_1\int\!\Id y_3\int\!\Id x_1\int\!\Id x_2\int\!\dvol(x_3)\cdot
\underbrace{\langle \delta s^\dagger(z_2),\delta s(z_{13})\rangle}_{-1}\cdot{}
\\
{}&\ \Bigl\{\Bigl\langle
\bigl(-\tfrac{\Id}{\Id z_2}\bigr) \bigl(\underbrace{\exp(q_{x_2})}_{x_2}\bigr)
\cdot \bigl( \underbrace{ {}^{\langle 10\rangle}\ %
\tfrac{\Id^2}{\Id z^2_{\boldsymbol{1}3}} (q^\dagger)+ 
{}^{\langle 9\rangle}\ %
\tfrac{\Id^2}{\Id y_1^2} (q^\dagger) }_{x_1} \bigr) \cdot
\tfrac{\Id^2}{\Id y_3^2 } (\underbrace{\cos q}_{x_3}) +{}
\\
{}&{}\qquad\quad 
+ \bigl(-\tfrac{\Id}{\Id z_2}\bigr)\bigl(\underbrace{\exp(q_{x_2})}_{x_2}\bigr)
\cdot \bigl(\underbrace{{}^{\langle 3\rangle}\ %
q^\dagger q_{x_1x_1} + {}^{\langle 4\rangle}\ %
\tfrac{\Id^2}{\Id y_1^2}(q^\dagger q)}_{x_1}\bigr) \cdot
\tfrac{\Id^2}{\Id y_3^2 } (\underbrace{-\sin q}_{x_3})
\Bigr\rangle \cdot
\underbrace{\langle \delta s(y_1),\delta s^\dagger(y_3) \rangle}_{+1} +{}
\\
{}&{}\quad
+\Bigl\langle
\bigl(-\tfrac{\Id}{\Id z_2}\bigr)\bigl(\underbrace{\exp(q_{x_2})}_{x_2}\bigr)
\cdot \bigl(\underbrace{{}^{\langle 13\rangle}\ %
q_{x_1x_1} + {}^{\langle 14\rangle}\ %
\tfrac{\Id^2}{\Id z_{\boldsymbol{1}3}^2}(q)}_{x_1}\bigr) \cdot
(\underbrace{ q^\dagger_{x_3x_3}\cdot (-\sin q)}_{x_3}) + {}
\\
{}&{}\qquad\quad 
+ \bigl(-\tfrac{\Id}{\Id z_2}\bigr) \bigl(\underbrace{\exp(q_{x_2})}_{x_2}\bigr)
\cdot \bigl(\underbrace{{}^{\langle 8\rangle}\ q q_{x_1x_1} }_{x_1}\bigr) \cdot
(\underbrace{ q^\dagger_{x_3x_3}\cdot (-\cos q)}_{x_3})
\Bigr\rangle \cdot 
\underbrace{\langle \delta s^\dagger(y_1),\delta s(y_3) \rangle}_{-1}
\Bigr\}.
\end{align*}
Each term $\langle1\rangle$\,--\,$\langle8\rangle$ meets its match 
from the other side of~\eqref{Jacobi4Schouten}, whereas
terms $\langle9\rangle$\,--\,$\langle14\rangle$ occur in pairs of opposite signs; therefore, they all cancel out in the
r.-h.s.\ of the Jacobi identity.
\end{example}

We conclude that within the true geometry of iterated variations, the Jacobi identity for~$\lshad\,,\,\rshad$~is
\begin{equation}\label{EqJacTrueRHS}
\Jac\bigl(F(x_1),G(x_2),H(x_3)\bigr)=0.
\end{equation}
By construction, its right\/-\/hand side is the functional whose density vanishes identically.

\section{The old 
approach to Jacobi's identity}\label{SecNaive}
\subsection{Cohomologically trivial r.-h.s.\ with nonzero density}
Let us ``forget'' the operational definition of variational Schouten bracket, that is, the way how the structure~$\lshad\,,\,\rshad$ is determined by the Batalin\/--\/Vilkovisky Laplacian, see Eq.~\eqref{EqDeviationDerivation}, 
now regarding the operation~$\lshad\,,\,\rshad$ as if it were introduced by formula~\eqref{EqFamiliar}. 
At first glance, the geometry becomes very simple:
all the intermediate objects are realised in the same way, namely, 
as integral functionals over the infinite jet superbundle~$J^\infty(\pi)$.
The price that one pays is the inconsistency of calculus~--- but let us postpone a counterexample till~\S\ref{SecCounter}.

\begin{example}\label{ExNaiveOneBase}
By taking the three functionals
\[
F=\int q^\dagger q q_{{xx}}\,\Id x,\qquad
G=\int q^\dagger_y \exp(q_x)\,\Id x,\qquad
H=\int q^\dagger_{{xx}} \cos q\,\Id x,
\]
and first, plugging any two of them into formula~\eqref{EqFamiliar}, one then inserts the output for an argument of the outer bracket in Jacobi's identity~\eqref{Jacobi4Schouten}. In this way one calculates the integral functional\footnote{The reader is invited to ponder whether it would be \emph{these} formulas that he or she is tempted to write.}
\[
\lshad G,H\rshad\quasieq\int \exp(q_x)\cdot \bigl\{
q^\dagger_x q_x^2 q_{xx} \cos q   + q^\dagger_{xx} q_x^2 \cos q 
+ q^\dagger_x q_{xx}^2 \sin q  \bigr\}\,\Id x,
\]
then making a short break. Resuming the job, one deduces that
\begin{align*}
\lshad F,\lshad G,&H\rshad\rshad\quasieq-\int\exp(q_x)\cdot\Bigl\{
 q^\dagger_x q q_{xx}^4 \sin q  + 2 q^\dagger_x q_x^4 q_{xx} \sin q 
     + 2 q^\dagger q_x^3 q_{xx}^2 \sin q   
\\
&     + 4 q^\dagger_{xx} q q_{xx}^3 \sin q  + 3 q^\dagger_x q q_{xx}^3 \cos q
     + 2 q^\dagger_x q_x q_{xx}^3 \sin q  - 2 q^\dagger_x q_x^2 q_{xx}^2 \cos q
\\
&     - 2 q^\dagger q_x q_{xx}^3 \cos q + 10 q^\dagger q_x^2 q_{xx}^2 \sin q  
     + 4 q^\dagger q_{xx}^2 q_{xxx} sin q  + 2 q^\dagger_{xxx} q q_{xx}^2 \sin q 
\\
&     - 4 q^\dagger_{xx} q q_{xx}^2 \cos q + q^\dagger_{xx} q q_x^4 \cos q 
     - q^\dagger_x q q_x^3 q_{xx}^2 \sin q  - q^\dagger_x  q q_x^4 q_{xx} \cos q
\\
&+ 3 q^\dagger_x q q_x q_{xx}^3 \cos q - 6 q^\dagger_x q q_x^2 q_{xx}^2 \sin q 
+ 4 q^\dagger_x q q_{xx}^2 q_{xxx} \sin q  + q^\dagger_{xx} q q_x q_{xx}^2\cos q 
\\
&+ 6 q^\dagger_{xx} q q_x^2 q_{xx} \sin q  + 6 q^\dagger_{xx} q q_{xx} q_{xxx} \sin q 
     - 2 q^\dagger_{xx} q q_x q_{xxx} \cos q + 4 q^\dagger_x q_x q_{xx}q_{xxx} \sin q 
\\
&     - 4 q^\dagger q_x q_{xx} q_{xxx} \cos q - 2 q^\dagger_{xxx} q q_x q_{xx}  \cos q
     + 2 q^\dagger_x q q_{xx} q_{4x} \sin q  + 2 q^\dagger_x q_x^5 \cos q
\\
&     + 2 q^\dagger q_{xx}^4 \sin q  - 4 q^\dagger q_{xx}^3 \cos q 
     - 4 q^\dagger_x q_x^2 q_{xxx} \cos q  + 10 q^\dagger_x q_x^3 q_{xx} \sin q 
\\
&\phantom{MMMMMMM}
+ 2  q^\dagger q_x^4 q_{xx} \cos q 
     - 4 q^\dagger_x q_x q_{xx}^2 \cos q + 4 q^\dagger_x q q_x q_{xx} q_{xxx} \cos q 
\Bigr\}\,\Id x.
\end{align*}
Likewise,
\[
\lshad F,G\rshad\quasieq\int q_{xx} \exp(q_x)\cdot \bigl\{
q^\dagger_x  q q_{xx} - 2 q^\dagger_x q_x - 2 q^\dagger q_{xx}
\bigr\}\,\Id x;
\]
here one stops for a while. The line of reasoning continues with
\begin{align*}
\lshad \lshad F,G\rshad,&H\rshad\quasieq -\int\exp(q_x)\cdot\Bigl\{
q^\dagger_x q q_x^2 q_{xx}^3 \cos q + 3 q^\dagger_{xx} q q_x^2 q_{xx}^2 \cos q 
   + 4 q^\dagger_{xx} q q_x^2 q_{xxx} \cos q 
\\
&   + 2 q^\dagger_{xxx} q q_x^2 q_{xx} \cos q + 2 q^\dagger_x q q_x^2 q_{4x} \cos q
   - 2 q^\dagger_{xxx} q_x^3 \cos q - 3 q^\dagger_x q_{xx}^3 \sin q 
\\
&   - 2 q^\dagger_{xx} q_x q_{xxx} \sin q  - 2 q^\dagger_{xxx} q_x q_{xx} \sin q 
   - 4 q^\dagger q_{xx} q_{4x} \sin q  + 3 q^\dagger_x q_x^3 q_{xx}^2 \cos q
\\
&   - 2 q^\dagger q_x^2 q_{xx}^3 \cos q + 4 q^\dagger_x q_x^3 q_{xxx} \cos q
   + 2 q^\dagger_{xx} q_x^3 q_{xx} \cos q - 4 q^\dagger q_x^2 q_{4x} \cos q
\\
&   + q^\dagger_{xx} q_x q_{xx}^2 \sin q  + q^\dagger_x q q_{xx}^4 \sin q 
   + 4 q^\dagger_{xx} q q_{xx}^3 \sin q  + 3 q^\dagger_x q_x q_{xx}^3 \sin q 
\\
&   - 3 q^\dagger_x q_x^2 q_{xx}^2 \cos q - 8 q^\dagger q_{xx}^2 q_{xxx} \sin q 
   + 2 q^\dagger_{xxx} q q_{xx}^2 \sin q  + 4 q^\dagger_x q q_{xx}^2 q_{xxx} \sin q 
\\
&   + 6 q^\dagger_{xx} q q_{xx} q_{xxx} \sin q  + 4 q^\dagger_x q_x q_{xx} q_{xxx} \sin q 
   + 2 q^\dagger_x q q_{xx} q_{4x} \sin q  - 2 q^\dagger q_{xx}^4 \sin q 
\\
&   - 6 q^\dagger_{xx} q_{xx}^2 \sin q  - 6 q^\dagger_{xx} q_x^2 q_{xx} \cos q
   - 8 q^\dagger_x q_x^2 q_{xxx} \cos q - 8 q^\dagger_x q_{xx} q_{xxx} \sin q 
\\
&\phantom{MMMMMMM}
- 8 q^\dagger q_x^2 q_{xx} q_{xxx} \cos q 
   + 4 q^\dagger_x q q_x^2 q_{xx} q_{xxx} \cos q 
\Bigr\}\,\Id x.
\end{align*}
Thirdly,
\[
\lshad F,H\rshad\quasieq -\int \bigl\{
q^\dagger_{xx} q q_x^2 \cos q  + 2 q^\dagger_x q_x^3 \cos q  
+ 2 q^\dagger q_x^2 q_{xx} \cos q 
+ 2 q^\dagger_x q_x q_{xx} \sin q  + 2 q^\dagger q_{xx}^2 \sin q 
\bigr\}\,\Id x,
\]
and now is the time for a pause. Finally, one obtains
\begin{align*}
\lshad G,\lshad F,&H\rshad \rshad\quasieq \int\exp(q_x)\cdot\Bigl\{
-q^\dagger_x q q_x^4 q_{xx} \cos q  - q^\dagger_{xx} q q_x^4 \cos q
    - 5 q^\dagger_x q q_x^2 q_{xx}^2 \sin q  
\\
&    + 2 q^\dagger_x q q_x q_{xx} q_{xxx} \cos q - 6 q^\dagger_{xx} q q_x^2 q_{xx} \sin q 
    + 2 q^\dagger_x q q_{xx}^3 \cos q - q^\dagger_x q_x^2 q_{xx}^2 \cos q
\\
&    + 2 q^\dagger_x q_x^3 q_{xx} \sin q  + 12 q^\dagger q_x^2 q_{xx}^2 \sin q 
    + 2 q^\dagger_{xx} q q_x q_{xxx} \cos q- 2 q^\dagger_x q_x q_{xx} q_{xxx} \sin q 
\\
&    - 8 q^\dagger q_x q_{xx} q_{xxx} \cos q + 2 q^\dagger_{xxx} q q_x q_{xx} \cos q
    + 4 q^\dagger_{xx} q q_{xx}^2 \cos q - 6 q^\dagger_x q_x q_{xx}^2 \cos q
\\
&    - 6 q^\dagger q_{xx}^3 \cos q - 2 q^\dagger_{xx} q_x q_{xxx} \sin q 
    - 8 q^\dagger_x q_{xx} q_{xxx} \sin q  - 2 q^\dagger_{xxx} q_x q_{xx} \sin q 
\\
&\phantom{MMMMMMM}
+ 2 q^\dagger q_x^4 q_{xx} \cos q
    - 4 q^\dagger q_{xx} q_{4x} \sin q  - 6 q^\dagger_{xx} q_{xx}^2 \sin q 
\Bigr\}\,\Id x.
\end{align*}
By using the software \textsf{Jets}~\cite{Jets} for symbolic calculations, one verifies that
\begin{equation}\label{EqLHSminusRHS}
\lshad F,\lshad G,H\rshad\rshad - \bigl(\lshad \lshad F,G\rshad,H\rshad + 
(-)^{(|F|-1)(|G|-1)}\lshad G,\lshad F,H\rshad \rshad\bigr) \cong 0.
\end{equation}
\end{example}

\subsection{Monitoring each argument's ``contribution'' to the r.-h.s.}
It is easy to explore the structure of 
cohomologically trivial functional in the right\/-\/hand side of the equi\-va\-len\-ce
\begin{equation}\label{EqJacRHSExact}
\Jac\bigl(F(x),G(x),H(x)\bigr)\cong0;
\end{equation}
for $F$,\ $G$,\ and~$H$ as above, we have that
\begin{align}
&\Jac\bigl(F(x),G(x),H(x)\bigr)=-\int\Id x\,\frac{\Id}{\Id x}\Bigl(\exp(q_x)\cdot
\Bigl\{ 
 2 q^\dagger q_{xx}^2  q_x^2  \cos q
- q^\dagger_x   q q_{xx}^2  q_x^2  \cos q
- 2 q^\dagger_x   q q_{xx} q_x^3  \sin q \notag
\\
&{}+ 5 q^\dagger_x  q q_{xx}^2  q_x \cos q
- 2 q^\dagger_x q_{xx} q_x^3  \cos q 
+ 4 q^\dagger q_{xx} q_x^3  \sin q  
- 2 q^\dagger_{xx}   q q_{xx} q_x^2  \cos q \notag
+ 4 q^\dagger  q_{xxx} q_x^2  \cos q
\\
&{}
+ 2 q^\dagger_x q_x^4  \sin q  
- 2 q^\dagger_x   q q_x^2 q_{xxx} \cos q 
+ 4 q^\dagger q_{xx}^3  \sin q  
- 10 q^\dagger q_{xx}^2  q_x \cos q 
- q^\dagger_x q_{xx}^2  q_x \sin q  
+ 2 q^\dagger_{xx} q_x^3  \cos q 
\Bigr\}\Bigr).\label{EqUnderDxRHS}
\end{align}
To track where these extra terms in the right\/-\/hand side came from, let us notice the following. First, the integrations by parts were always attached to the respective vertical differentials in the fibres of~$J^\infty(\pi)$. 
Consequently, whenever two variations fell on (whatever remained of) one functional, they produced the terms with operators 
such that the total and partial derivatives are interlaced, 
e.g., in 
\begin{equation}\label{EqCombineWrong}
\Bigr(-\frac{\overrightarrow{\Id}}{\Id\bx}\Bigr)^{\tau} \circ
\frac{\overrightarrow{\dd}}{\dd\bq_\tau} \circ
\Bigr(-\frac{\overrightarrow{\Id}}{\Id\bx}\Bigr)^{\sigma} \circ
\frac{\overrightarrow{\dd}}{\dd\bq^\dagger_\sigma}
\end{equation}
or in a similar sequence of interlacing total and partial derivatives~--- 
with any other combinations of the variables~$\bq$ and~$\bq^\dagger$.

Which is worse, the construction of nested brackets in a term like 
$\lshad F,\lshad G,H\rshad\rshad$ according to formula~\eqref{EqFamiliar} prescribes that, in the course of integration by parts in the bracket of~$F$ and entire~$\lshad G,H\rshad$, whenever a partial derivative~$\dd/\dd\bq_\tau$ or~$\dd/\dd\bq^\dagger_\tau$ with~$|\tau|>0$ falls on the density of~$H$, the total derivative~$(-\vec{\Id}/\Id\bx)^\tau$ does act on~$G$ because it  spreads over~$G$ and~$H$ via Newton's binomial formula. Note the 
cancellation of 
terms in which the total derivative~$(-\vec{\Id}/\Id\bx)^\tau$ falls exclusively on the image of its ``native'' partial derivative~$\dd/\dd\bq_\tau$ or~$\dd/\dd\bq^\dagger_\tau$; this was proved in~\cite{sqs13} and illustrated in~\S\ref{SecTrue} above.\footnote{An explanation why the mess of redundant cross\/-\/terms remains overall exact is an instructive exercise for the reader.}

\begin{example}\label{ExNaimeManyBases}
Let us analyse how the remaining cross\/-\/terms emerge and contribute to the right\/-\/hand side of~\eqref{EqJacRHSExact}. To visualize their origin from one of the three functionals, we formally denote by~$x$,\ $y$,\ and~$z$ the respective base variables so that
\[
F=\smash{\int q^\dagger q q_{{xx}}}\,\Id x,\qquad
G=\smash{\int q^\dagger_y \exp(q_y)}\,\Id y,\qquad
H=\smash{\int q^\dagger_{{zz}} \cos q}\,\Id z;
\]
the restriction $\Jac\bigl(F(x),G(y),H(z)\bigr)\bigr|_{x=y=z}$ 
to the diagonal 
at the end of the day will yield~\eqref{EqUnderDxRHS} in the right\/-\/hand side of~\eqref{EqJacRHSExact}.

We emphasize that the genuine contribution to the right\/-\/hand side 
of~\eqref{EqLHSminusRHS} 
is identically zero; the rest of its density is \emph{naught} (i.e., not a mathematical description of any existing object) which equals
\begin{align*}
\exp&(q_y)\cdot\smash{\Bigl\{ }
 2 q^\dagger q_{xx} q_y q_{yy} q_{zz} \cos q 
- 8 q^\dagger q_{xx} q_y q_z q_{yz} \sin q  
+ q^\dagger_{xx} q q_y q_{yy} q_{zz} \cos q 
\\ &
- 4 q^\dagger_{xx} q q_y q_z q_{yz}\sin q  
+ 2 q^\dagger_x q_x q_y q_{yy} q_{zz} \cos q 
- 8 q^\dagger_x q_x q_y q_z q_{yz} \sin q  
+ q^\dagger_y q q_{xx} q_y^2  q_z^2 \cos q 
\\ & 
+ q^\dagger_y q q_{xx} q_{yy} q_z^2 \sin q   
- q^\dagger_y q q_{xx} q_{yy} q_{yz}^2 \sin q   
+ q^\dagger_y q q_{xx} q_y^2  q_{zz} \sin q  
+ q^\dagger_{yy} q q_{xx} q_y q_z^2 \sin q   
\end{align*}
\begin{align*}
&
- 2 q^\dagger_{yz} q q_{xx} q_{yy} q_{yz} \sin q  
- 2 q^\dagger_y q q_{xx} q_{yy} q_{yzz} \sin q  
+ q^\dagger_{zz} q q_{xx} q_y q_{yy} \cos q 
- q^\dagger_y q q_{xx} q_{yy} q_{zz} \cos q 
\\ &
- 2 q^\dagger_y q q_{xx} q_{yz} q_{yyz} \sin q  
- q^\dagger_{yy} q q_{xx} q_y q_{zz} \cos q 
- 2 q^\dagger_y q q_{xx} q_y q_{yzz} \cos q 
- 2 q^\dagger_{yy} q q_{xx} q_{yz} q_z \cos q 
\\ &
- 2 q^\dagger_y q q_{xx} q_{yyz} q_z \cos q 
+ q^\dagger_y  q q_{xy}^2 q_{yy} q_z^2 \cos q 
+ q^\dagger_y q_{xx} q_y q_{yy} q_z^2  \cos q 
+ 2 q^\dagger_y q_x q_{xy} q_{yy} q_z^2  \cos q 
\\ &
+ q^\dagger_y  q q_{xy}^2 q_{yy} q_{zz} \sin q  
+ 2 q^\dagger_y  q q_{xxy} q_{yy} q_z^2 \cos q 
+ 2 q^\dagger_{xy} q q_{xy} q_{yy} q_z^2  \cos q 
+ 2 q^\dagger_y q q_{xy} q_{xyy} q_z^2  \cos q 
\\ &
+ q^\dagger_y q_{xx} q_y q_{yy} q_{zz}  \sin q  
+ 2 q^\dagger_y q_x q_{xy} q_{yy} q_{zz}  \sin q  
+ 2 q^\dagger_y q q_{xxy} q_{yy} q_{zz} \sin q  
+ 2 q^\dagger_{xy} q q_{xy} q_{yy} q_{zz} \sin q  
\\ &
+ 2 q^\dagger_y q q_{xy} q_{xyy} q_{zz}  \sin q  
- 8 q^\dagger q_x q_{yy} q_{xz} q_z \sin q  
+ q^\dagger_y q q_x^2 q_{yy} q_z^2  \cos q 
+ q^\dagger_y q q_x^2 q_{yy} q_{zz} \sin q  
\\ &
- 2 q^\dagger_y q q_x q_{yy} q_{xzz} \cos q 
+ 4 q^\dagger_{yy} q q_x q_{xz} q_z \sin q  
- 2 q^\dagger q_{xx} q_y q_{yy} q_z^2 \sin q   
- q^\dagger_{xx} q q_y q_{yy} q_z^2  \sin q  
\\ &
- 2 q^\dagger_x q_x q_y q_{yy} q_z^2 \sin q  
- 4 q^\dagger_{xy} q_{xy} q_{zz} \sin q  
- 2 q^\dagger_{zz} q_x^2 q_{yy} \cos q 
+ 4 q^\dagger q_{yy} q_{xz}^2 \cos q 
\\ &
+ 4 q^\dagger_x q_{yy} q_{xzz} \sin q  
+ 2 q^\dagger_{xxz} q_{yy} q_z \sin q  
+ 4 q^\dagger_{xz} q_{yy} q_{xz} \sin q  
+ 4 q^\dagger_z q_{yy} q_{xxz} \sin q  
\\ &
+ 4 q^\dagger q_{yy} q_{xxzz} \sin q  
- 2 q^\dagger_{zz} q_{xxy} q_y \sin q  
+ q^\dagger_{yy} q_{xx} q_z^2  \cos q 
- 2 q^\dagger_{yy} q q_{xz}^2  \cos q 
\\ &
+ 2 q^\dagger_{yy} q_{xxz} q_z \sin q  
+ 4 q^\dagger q_{xx} q_{yz}^2  \cos q 
+ 2 q^\dagger_{xx} q q_{yz}^2  \cos q 
+ 4 q^\dagger_x q_x q_{yz}^2  \cos q 
\\ &
- q^\dagger_{xx} q_{yy} q_z^2  \cos q 
- 4 q^\dagger_x q_{xyy} q_z^2 \cos q 
- 2 q^\dagger_{xxy} q_y q_z^2  \cos q 
- 4 q^\dagger_y q_{xxy} q_z^2   \cos q 
\\ &
- 4 q^\dagger q_{xxyy} q_z^2  \cos q 
- 4 q^\dagger_{xy}  q_{xy} q_z^2  \cos q 
- 4 q^\dagger_x q_{xyy} q_{zz} \sin q  
- 2 q^\dagger_{xxy} q_y q_{zz}  \sin q 
\\ &
- 4 q^\dagger_y q_{xxy} q_{zz} \sin q  
- 4 q^\dagger q_{xxyy} q_{zz} \sin q  
+ 4 q^\dagger q_{xxz} q_{yy} q_z \cos q 
+ q^\dagger_{zz} q q_{xy}^2 q_{yy} \sin q  
\\ &
- q^\dagger_{zz} q_{xx} q_y q_{yy} \sin q  
+ 2 q^\dagger_{zz} q q_{xy} q_{xyy} \sin q  
+ q^\dagger_{yy} q q_x^2 q_z^2 \cos q 
+ q^\dagger_{yy} q q_x^2 q_{zz} \sin q  
\\ &
- 2 q^\dagger_y q q_{yy} q_{xz}^2 \cos q 
- 2 q^\dagger_{yy} q q_x q_{xzz} \cos q 
+ 2 q^\dagger_y q_{xxz} q_{yy} q_z  \sin q  
- 4 q^\dagger_x q_x q_{yz} q_{yyz} \sin q  
\\ &
+ 4 q^\dagger_x q_x q_y q_{yzz} \cos q 
- q^\dagger_{zz} q q_{xx} q_y^2 \sin q  
- q^\dagger_{yy} q q_{xx} q_{yz}^2  \sin q  
- 2 q^\dagger_y q q_{xx} q_{yz}^2  \cos q 
\\ &
- 2 q^\dagger_{yyz} q q_{xx} q_{yz} \sin q  
- 2 q^\dagger_{yy} q q_{xx} q_{yzz} \sin q  
- 2 q^\dagger_{yz} q q_{xx} q_{yyz} \sin q  
- 2 q^\dagger_y q q_{xx} q_{yyzz} \sin q  
\\ &
+ 2 q^\dagger_{yzz} q q_{xx} q_y \cos q 
- 2 q^\dagger q_{yy} q_{xy}^2 q_z^2 \cos q 
+ q^\dagger_{yy} q q_{xy}^2 q_z^2  \cos q 
+ 2 q^\dagger_y q_{xx} q_{yy} q_z^2  \cos q 
\\ &
+ q^\dagger_{yy} q_{xx} q_y q_z^2  \cos q 
- q^\dagger_{xx} q_y q_{yy} q_z^2  \cos q
- 4 q^\dagger_x q_{xy} q_{yy} q_z^2  \cos q 
+ 2 q^\dagger_{yy} q_{xy} q_x q_z^2 \cos q 
\\ &
+ 2 q^\dagger_y q_x q_{xyy} q_z^2  \cos q 
- 2 q^\dagger q_{xy}^2 q_{yy} q_{zz}  \sin q  
- 4 q^\dagger q_{xxy} q_{yy} q_z^2  \cos q 
+ q^\dagger_{yy} q q_{xy}^2 q_{zz} \sin q  
\\ &
+ 2 q^\dagger_y q_{xxy} q_y q_z^2  \cos q 
+ 2 q^\dagger_y q_x q_{xyy} q_{zz} \sin q  
- 4 q^\dagger q_{xxy} q_{yy} q_{zz} \sin q  
+ 2 q^\dagger_y q_{xxy} q_y q_{zz} \sin q  
\\ &
+ 2 q^\dagger_{yy} q q_{xxy} q_{zz} \sin q  
+ 2 q^\dagger_y q q_{xxyy} q_{zz} \sin q  
+ 2 q^\dagger_{xyy} q q_{xy} q_{zz} \sin q 
+ 2 q^\dagger_{xy} q q_{xyy} q_{zz} \sin q  
\\ &
- 4 q^\dagger q_{xy} q_{xyy} q_{zz} \sin q 
- 2 q^\dagger q_x^2 q_{yy} q_z^2  \cos q 
- 2 q^\dagger q_{xx} q_{yy} q_z^2  \sin q  
+ 2 q^\dagger_{yy} q q_{xxy} q_z^2 \cos q 
\\ &
+ 2 q^\dagger_y q q_{xxyy} q_z^2 \cos q 
+ 2 q^\dagger_{xyy} q q_{xy} q_z^2 \cos q
+ 2 q^\dagger_{xy} q q_{xyy} q_z^2 \cos q 
- 4 q^\dagger q_{xy} q_{xyy} q_z^2 \cos q 
\\ &
+ q^\dagger_y q_{xx} q_{yy} q_{zz} \sin q 
+ q^\dagger_{yy} q_{xx} q_y q_{zz} \sin q  
- q^\dagger_{xx} q_{yy} q_y q_{zz}  \sin q 
- 4 q^\dagger_x q_{xy} q_{yy} q_{zz}  \sin q 
\\ &
+ 2 q^\dagger_{yy} q_x q_{xy} q_{zz}  \sin q 
- 2 q^\dagger q_{xx} q_y^2 q_z^2  \cos q 
- q^\dagger_{xx}  q q_y^2 q_z^2  \cos q
- 2 q^\dagger_x q_x q_y^2 q_z^2  \cos q
\\ &
- 2 q^\dagger q_{xx} q_{yy} q_{yz}^2  \sin q  
- 2 q^\dagger q_{xx} q_y^2 q_{zz} \sin q  
- q^\dagger_{xx} q q_{yy} q_{yz}^2 \sin q  
- q^\dagger_{xx} q q_y^2 q_{zz} \sin q  
\\ &
- 2 q^\dagger_x q_x q_{yy} q_{yz}^2  \sin q  
- 2 q^\dagger_x q_x q_y^2  q_{zz} \sin q  
- 4 q^\dagger q_{xx} q_{yz} q_{yyz} \sin q 
+ 4 q^\dagger q_{xx} q_y q_{yzz}  \cos q
\\ &
- 2 q^\dagger_{xx} q q_{yz} q_{yyz} \sin q  
+ 2 q^\dagger_{xx} q q_y q_{yzz} \cos q
- 2 q^\dagger_x q_x q_{yy} q_z^2   \sin q 
- 2 q^\dagger q_x^2 q_{yy} q_{zz}  \sin q  
\\ &
+ q^\dagger_{zz} q q_x^2 q_{yy} \sin q  
+ 2 q^\dagger q_{xx} q_{yy} q_{zz}  \cos q
+ 2 q^\dagger_x q_x q_{yy} q_{zz}  \cos q
+ 4 q^\dagger_x q_{xz} q_{yy} q_z \cos q
\\ &
+ 4 q^\dagger q_x q_{yy} q_{xzz}  \cos q
- 2 q^\dagger_{xzz} q q_x q_{yy} \cos q 
+ q^\dagger_y q q_{xx} q_y q_{yy} q_z^2   \sin q  
- q^\dagger_y q q_{xx} q_y q_{yy} q_{zz}  \cos q
\\ &
- 2 q^\dagger_y q q_{xx} q_{yy} q_{yz} q_z  \cos q
+ 4 q^\dagger_y q q_{xx} q_y q_{yz} q_z  \sin q 
+ 4 q^\dagger_y q q_x q_{xz} q_{yy} q_z \sin q 
\smash{\Bigr\}}.
\end{align*}
Clearly, it would have been fairly impossible to calculate this quantity without suitable software~(\cite{Jets,SsTools}). Yet it is this ephemeral fiction that consumed most of the processor time; the genuine right\/-\/hand side (which there is none) was calculated by hand, see Eq.~\eqref{EqJacTrueRHS} on p.~\pageref{EqJacTrueRHS}.
\end{example}

\section{A yet another manifestation of old formalism's inconsistency%
}\label{SecCounter}
%
We agree that the presence of exact terms in the right\/-\/hand side of~\eqref{EqJacRHSExact} does not discredit any result known from the theories in which all objects are processed only by the Schouten bracket (possibly, in its transcript~\eqref{EqFamiliar}). For instance, the claims remain true for variational Poisson bi\/-\/vectors, the Poisson cohomology groups they give rise to (see~\cite{KacDeSoleJ}), or bi\/-\/Hamiltonian cohomology~\cite{DubrZhang2001}. 
The contrast between~\eqref{EqCombineTrue} and~\eqref{EqCombineWrong} makes no harm in that narrow sub\/-\/class of problems which are posed in the frames of variational symplectic supergeometry of parity\/-\/even variables~$\bq$ and their parity\/-\/odd canonical conjugates~$\bq^\dagger$ over~$\bx\in M^m$.

Unfortunately, there is much amiss if the superbundle~$\pi$ stays the same but the class of problems to\/-\/consider is less narrow 
(e.g., see~\cite{
gvbv,dq15,cycle14} and references therein). 
Namely, the Batalin\/--\/Vilkovisky Laplacian~$\Delta$ stops being a graded derivation of the Schouten bracket,
\begin{equation}\label{EqZimes}
\Delta\bigl(\lshad F,G\rshad\bigr) = \lshad\Delta F,G\rshad 
+\smash{(-)^{|F|-1}}\lshad F,\Delta G\rshad,
\end{equation}
whence the parity\/-\/odd linear operator~$\Delta$ stops being a 
differential,\footnote{Equally bad an option would it be to postulate the validity of main relations between the Batalin\/--\/Vilkovisky Laplacian~$\Delta$ and variational Schouten bracket~$\lshad\,,\,\rshad$, so that neither their proof nor explicit examples are possible any longer.
}
violating~\eqref{EqLaplace2Zero}. Let us substantiate this claim 
by 
a counterexample.

It is often accepted that the Batalin\/--\/Vilkovisky Laplacian~$\Delta$ is no more than the parity\/-\/odd linear operator which acts on a given integral functional $H=\int h\bigl(\bx,[\bq],[\bq^\dagger]\bigr)\,\dvol(\bx)$ by the formula\footnote{%
The genuine linear operation~$\Delta$ is extended to the vector space of formal products of integral functionals by the Leibniz rule for~$\vec{\dd}/\dd\bq_\tau$ and~$\vec{\dd}/\dd\bq^\dagger_\sigma$, yielding the operational definition of variational Schouten bracket, see~\eqref{EqDeviationDerivation}.}
\begin{multline}\label{EqBVNaive}
\Delta(H) \quasieq
\int \frac{\vec{\delta}}{\delta\bq}\cdot\frac{\vec{\delta}}{\delta\bq^\dagger}\,
(h)\,\bigl(\bx,[\bq],[\bq^\dagger]\bigr)\,\dvol(\bx)={}\\
{}=\int \sum_{i=1}^m \sum_{\substack{|\sigma|\geq0 \\ |\tau|\geq0}}
\biggl(
\Bigr(-\frac{\overrightarrow{\Id}}{\Id\bx}\Bigr)^{\tau} \circ
\frac{\overrightarrow{\dd}}{\dd q^i_\tau} \circ
\Bigr(-\frac{\overrightarrow{\Id}}{\Id\bx}\Bigr)^{\sigma} \circ
\frac{\overrightarrow{\dd}}{\dd q^\dagger_{i,\sigma}}
\biggr)\ %
h (\bx,[\bq],[\bq^\dagger])\,\dvol(\bx).
\end{multline}
We now
demonstrate\footnote{Let us warn the reader that all claims in Counterexample~\ref{ExCounter} about any equalities between functionals (specifically, for $\Delta F$ and~$\Delta H$ or for $\lshad F,H\rshad$ and~$\Delta\bigl(\lshad F,H\rshad\bigr)$, etc.) should be viewed as the classical parable about a cage which contains an elephant but carries an inscription ``\textsc{Buffalo}''~--- one may not trust his own eyes. We refer to Remark~\ref{RemSynonyms} below and also to Example~2.4 
in~\cite
{gvbv},
in which we explicitly calculate the objects $\Delta F$ and~$\Delta H$ 
or~$\Delta\bigl(\lshad F,H\rshad\bigr)$, 
confirming that equality~\eqref{EqNotHolds} \emph{is} valid.} 
that formula~\eqref{EqBVNaive} is oversimplified to the extent that it is not able to let 
the Batalin\/--\/Vilkovisky Laplacian 
satisfy important identity~\eqref{EqZimes}.

\begin{counterEx}\label{ExCounter}
Let us denote by~$f$ and~$h$ the respective integrands in 
$F = \int q^\dagger q q_{xx}\,\Id x$ and $H = \int q^\dagger_{xx}\cos q\,\Id x$.
One eagerly calculates\footnote{Note that since all the four variational derivatives contain at most one parity\/-\/odd $q^\dagger$ or its derivatives, the directions of all the derivations 
do not actually matter~--- i.e., reversing their direction would not result in minus signs.}
\begin{align*}
\frac{\delta f}{\delta q} 
 &=  q^\dagger q_{xx} + \tfrac{\Id^2}{\Id x^2}\bigl(q^\dagger q\bigr)
 = 2q^\dagger q_{xx} + 2q^\dagger_x q_x + q^\dagger_{xx} q, &
\frac{\delta f}{\delta q^\dagger} &= qq_{xx}, \\
\frac{\delta h}{\delta q^\dagger} 
 &=\tfrac{\Id^2}{\Id x^2}\bigl(\cos q\bigr) 
 = -q_{xx}\sin q  - q_x^2\cos q, &
\frac{\delta h}{\delta q} &= -q^\dagger_{xx}\sin q.
\end{align*}
Consider first $\Delta\bigl(\lshad F,H\rshad\bigr)$. 
Our new working formula~\eqref{EqBVNaive}, combined with the primary exercise
$\delta/\delta q\circ\Id/\Id x \equiv 0
\equiv \delta/\delta q^\dagger\circ\Id/\Id x$,
suggests that one writes $\Delta(G)\quasicong\int\bigl(\partial/\partial q \circ\partial/\partial q^\dagger\bigr)(g)\,\Id x$ for any integral functional $G=\int g\,\Id x$. This implies that only those terms survive under~$\Delta$
in which the density of~$\lshad F,H\rshad$ carries 
$q^\dagger$ without derivatives with respect to~$x$. 
So, one takes into account only
\[
\lshad F,H \rshad \quasieq \int\bigl(
(2q^\dagger q_{xx} + \ldots) \cdot (-q_{xx}\sin q  - q_x^2\cos q) + \ldots
\bigr)\,\Id x,
\]
where the dots indicate the irrelevant terms with~$q^\dagger_x$ or~$q^\dagger_{xx}$. 
This produces the functional 
\begin{multline*}
\Delta\bigl(\lshad F,H\rshad\bigr)
\quasicong -2\int\frac\partial{\partial q}\frac\partial{\partial q^\dagger}
 \bigl(q^\dagger q_{xx}^2\sin q  + q^\dagger q_x^2 q_{xx}\cos q\bigr)\,\Id x 
\\
= -2\int\frac\partial{\partial q}\bigl(q_{xx}^2\sin q  + q_x^2 q_{xx}\cos q\bigr)\,\Id x
 = -2\int\bigl(q_{xx}^2\cos q - q_x^2 q_{xx}\sin q \bigr)\,\Id x,
\end{multline*}
the integrand of which is not cohomogically trivial (as could be readily seen by calculating its variational derivative, which gives nonzero).

On the other hand, $\Delta F\quasicong\int q_{xx}\,\Id x \cong 0$,
whence $\lshad \Delta F,H\rshad\quasieq\lshad\, 0,H\rshad=0$.
At the same time, the density~$h$ contains no $q^\dagger$ 
but only the second derivative~$q^\dagger_{xx}$, 
which means now that~$\Delta H\quasicong0$ as well, 
whence $\lshad F,\Delta H\rshad\quasieq\lshad F,0\,\rshad=0$.
In conclusion to this ``counterexample,''
\begin{equation}\label{EqNotHolds}
\Delta\bigl(\lshad F,H\rshad\bigr) \quasineq
\lshad \Delta F,H\rshad + (-)^{|F|-1}\lshad F, \Delta H\rshad = 0.
\end{equation}
This ``contradiction'' marks the limits of jet\/-\/bundle approach~\cite{KuperCotangent} to the BV-\/geometry via Vinogradov's $\cC$-\/spectral sequence $E^{p,q}_i$ (specifically, by using the upper line $E^{n,q}_1$ of its first term such that $E^{n,0}_1 = \bar{H}{}^n(\pi) \ni F,G,H$, cf.~\cite{
VinogradovCSpecII}). 
The resolution of apparent difficulties is achi\-e\-ved in~\cite{gvbv,sqs13}, 
where we rigorously prove the validity of identities~\eqref{EqZimes}, \eqref{EqDeviationDerivation}, and
\begin{equation}\label{EqLaplace2Zero}
\Delta^2=0.
\end{equation}
In particular, identity~\eqref{EqZimes} does of course hold 
for the functionals~$F$ and~$H$ as above~--- this is confirmed in~\cite[pp.\,34--36]{gvbv}.
\end{counterEx}

\begin{rem}\label{RemSynonyms}
A very interesting effect which the true theory of iterated variations now offers
is the natural existence of synonyms for zero functional; that is, there are objects which would take every section~$\bolds\in\Gamma(\pi)$ of the superbundle to~$0\in\Bbbk$ but which, belonging in fact to spaces larger than that for the trivial cohomology class $\int \Id_h\eta 
\in E^{n,0}_1$, can contribute \emph{non}trivially to the output of a calculation. (To comprehend why this is possible, compare~\eqref{EqCombineTrue} with~\eqref{EqCombineWrong} and refer to~\cite[\S1.4]{gvbv} for more explanation.) Such is the object~$\Delta H$ in~\eqref{EqNotHolds} for the functional $H=\int q^\dagger_{xx}\,\cos q\,\Id x$, which we used in all the examples here.
\end{rem}

This phenomenon manifests the first main slogan in the geometry of iterated variations: \textbf{no calculation can be interrupted at any intermediate step}.

\begin{rem}
The second guiding principle is that 
{the integrations by parts always fall only on the functionals which they stem from but never hit the contributions from other functionals}. Therefore, the variation of~$G$ or~$H$ within~$\lshad G,H\rshad$ in a term like $\lshad F,\lshad G,H\rshad \rshad$ is such that the specific choice of~$H$ (or, respectively,~$G$) does not matter; the structure
~$\lshad\,,\,\rshad$ is uniquiely defined by~\eqref{EqUniqueDef} --\,or, equivalently, by~\eqref{EqDeviationDerivation}\,-- for all pairs of functionals.

Conversely, formula~\eqref{EqFamiliar} tells us that the bi\/-\/linear operation introduced by it makes the derivations which will fall on~$G$ or~$H$ within~$\lshad G,H\rshad$ in the future calculation of $\lshad F,\lshad G,H\rshad \rshad$
explicitly dependent on a choice of the other argument (i.e.,~$H$ or~$G$, respectively). In other words, formula~\eqref{EqFamiliar} encodes \emph{infinitely many} structures (roughly speaking, its own structure for each pair of arguments). This is in contrast to the standard 
idea${}^{\text{\ref{FootCommutator} on p.~\pageref{FootCommutator}}}$
that the variational Schouten bracket is a \emph{unique} extension of the commutator~$[\,,\,]$ of evolutionary vector fields on the jet bundle~$J^\infty(\pi)$ to the space of variational multi\/-\/vectors on~it.
\end{rem}

\ack
The author thanks the Organizing committee of XXIII International conference `Integrable systems \& quantum symmetries'
(22\,--\,28 June 2015; CVUT Prague, Czech Republic) for a warm atmosphere during the meeting. 
This research was partially supported by JBI~RUG project~103511 (Groningen). A~part of this work 
was done while the author was visiting at the~$\smash{\text{IH\'ES}}$ (Bures\/-\/sur\/-\/Yvette, France); the hospitality and partial financial support of that institution are gratefully acknowledged.

%

\medskip

\providecommand{\newblock}{}

\end{document}